# Acceleration and Parallelization Methods for ISRS EGN Model

Ruiyang Xia, Guanjun Gao, Zanshan Zhao, Haoyu Wang, Kun Wen, and Daobin Wang

*Abstract*—**The enhanced Gaussian noise (EGN) model, which accounts for inter-channel stimulated Raman scattering (ISRS), has been extensively utilized for evaluating nonlinear interference (NLI) within the C+L band. Compared to closed-form expressions and machine learning-based NLI evaluation models, it demonstrates broader applicability and its accuracy is not dependent on the support of large-scale datasets. However, its high computational complexity often results in lengthy computation times. Through analysis, the high-frequency oscillations of the four-wave mixing (FWM) efficiency factor integrand were identified as a primary factor limiting the computational speed of the ISRS EGN model. To address this issue, we propose an approximation method to derive a closed-form expression for the FWM efficiency factor, which provides both high accuracy and high computational efficiency. Numerical results demonstrate that the method proposed in this work could achieve low error levels under high ISRS influence levels, with an absolute mean error (MAE) of approximately 0.0033 dB. Furthermore, a parallel computation strategy targeting independent regions within the integration domain is proposed and optimized, which further significantly improves computing efficiency.**

## I. INTRODUCTION

As modern optical communication systems evolve toward ultra-wideband transmission, nonlinear effects in optical fibers have become increasingly complex and prominent[1-2]. For optical networks with large number of optical add-drop multiplexer (OADM) nodes and flexibly switched optical wavelength channels, accurate and time-efficient quality of transmission (QoT) estimation becomes extremely important[3-5]. However, brute-force numerical approaches, such as solving the nonlinear Schrödinger equation (NLSE) using the split-step Fourier method (SSFM), are no longer practical due to their high computational complexity[6-7]. The advancement of coherent optical communication with digital signal processing (DSP) has driven the development of the Gaussian noise (GN) model, which models nonlinear noise in uncompensated transmission (UT) as approximately Gaussian[8-11]. Although the GN model provides a relatively efficient method for nonlinear evaluation, it does not account for modulation format effects[12], and several studies have indicated that it may overestimate nonlinear interference (NLI)[13-16]. In 2013, Carena A. et al. proposed the enhanced Gaussian noise (EGN) model by introducing some additional correction terms into the GN model. The EGN model accounts for the modulation format dependence of NLI, thereby offering a more precise assessment of NLI[17]. However, the inter-channel power transfer caused by inter-channel stimulated Raman scattering (ISRS) was not considered in the [17], which cannot be neglected in C+L band transmission scenarios[18]. In 2018, Cantono M. et al. considered the simultaneous effect of ISRS and NLI generation, and introduced ISRS into the GN model [19]. In 2020, building upon the previous work, Rabbani H. et al. further introduced the ISRS into the EGN model [20]. In 2020, Rabbani H. et al. expanded upon the results [17] by incorporating the interaction between the Kerr effect and ISRS. Additionally, they extended the study to various wavelength division multiplexing (WDM) channel configurations with flexible modulation formats, variable symbol rates, heterogeneous fiber segments, and non-uniform power distributions. The scheme proposed in the [20] could provide more accurate NLI evaluation with broad applicability, but its computational complexity is significantly higher than that of the conventional GN model. Despite its relative speed advantage over SSFM. For scenarios that require real-time evaluation of transmission performance quality, e.g., digital twins, computation time remains a serious issue.

Moreover, many efforts have also been made to further reduce the computation time of the GN or EGN models, resulting in the development of closed-form expressions capable of producing results in near real-time [21]. In 2017, Semrau D. et al. first derived a closed-form expression for the GN model that accounts for ISRS effects[22]. In 2019, this initial work was extended to incorporate dispersion slope effects and to support arbitrary launch power distributions[23]. Subsequently, this team introduced the effects of cross-phase modulation (XPM), allowing for the consideration of various modulation formats[24].

This work was supported in part by National Key Research and Development Program of China (2022YFB2903303) and National Natural Science Foundation of China (No. 62141505, 61367007 and 62371064); (Corresponding author: Guanjun Gao; Daobin Wang).

Ruiyang Xia, Daobin Wang, and Kun Wen are with the School of Science, Lanzhou University of Technology, LanZhou 730050, China (e-mail: 232080901001@lut.edu.cn; cougarlz@lut.edu.cn; 1507996369@qq.com)

Guanjun Gao, Zanshan Zhao, and Haoyu Wang are with the School of Electronic Engineering, Beijing University of Posts and Telecommunications, Beijing 100876, China (e-mail: ggj@bupt.edu.cn; zzs@bupt.edu.cn; buptwhy@bupt.edu.cn)

To develop a real-time NLI evaluation model with sufficient accuracy, researchers have explored machine learning techniques, either to improve the accuracy of closed-form expressions or to construct data-driven versions of the EGN model. In 2020, Zefreh M. R. et al. utilized machine learning to refine the incoherent GN closed-form expression[25]. The optimized closed-form expression (CFM2) demonstrated accuracy comparable to the EGN model on the test set. In 2021, Müller J. et al. trained an artificial neural network (ANN) for NLI estimation[26]. In field measurement data, this model provided real-time results with a maximum signal-to-noise ratio (SNR) error of less than 0.5 dB. In 2022, Müller J. et al. used the self-channel interference (SCI) results calculated by the EGN model, along with relevant system parameters, as inputs to an Extreme Gradient Boosting (XGB) algorithm to obtain NLI predictions[27]. Since it only requires integration over a narrow frequency band, it is faster than using the full EGN model. [25-27] used the EGN model to generate data for machine learning applications and achieved effective results in specific optical fiber link configurations.

Although the aforementioned closed-form expressions and machine learning-based approaches offer better real-time performance, the ISRS EGN model is still widely used in many practical applications. On one hand, the ISRS EGN model is derived from the EGN model, which provides a complete and detailed integral-based modeling of NLI. It is capable of consistently achieving highly accurate results across a wide range of scenarios. On the other hand, the reliability of the ISRS EGN model does not depend on the scale or quality of training data and is not affected by challenges related to generalization[28-30]. To address the issue of slow computation speed in the ISRS EGN model, this work proposes an accurate and efficient acceleration method. Specifically, the contributions of this work are as follows:

(1) Based on the proposed piecewise approximation method, a closed-form expression for the four-wave mixing (FWM) efficiency factor considering the impact of ISRS is derived. The proposed formulation yields highly accurate results, even in scenarios with significant ISRS effects.

(2) A hybrid approximation method is proposed that significantly reduces computational complexity while maintaining high accuracy. Compared to the full integral form of the ISRS EGN model, this method achieves a computation time reduction of over 99.5%, with a mean absolute error (MAE) of approximately 0.0033 dB.

(3) A parallel computation method targeting independent regions within the integration domain of the ISRS EGN model is proposed, along with optimized strategies to achieve load balancing and efficient scheduling. This approach achieves a further substantial enhancement in computational efficiency.

The remainder of this paper is organized as follows. Section II first introduces the overall strategy, followed by detailed descriptions of the proposed parallel acceleration method and the closed-form FWM efficiency factor. Section III evaluates the performance of the proposed acceleration methods and provides discussions. Finally, Section IV concludes the paper.

## II. The Acceleration Methods

### *A. Overall Acceleration Schematic*

Fig. 1 illustrates the overall schematic of the proposed ISRS EGN model acceleration method. The base model utilized in this work was introduced in [20], with the relevant equations provided in Appendix B as Eqs. (21)-(30). This model accounts for the combined effects of SCI, cross-channel interference (XCI), multi-channel interference (MCI), and ISRS, enabling accurate NLI evaluation in heterogeneous fiber links over the C+L band. In Eq. (21), the model's integration domain is divided into multiple independent regions, also referred to as "islands", by parameters $\kappa_1$ , $\kappa_2$ , and $l$ , on which computations are performed independently. The first acceleration method proposed in this work implements parallel computation across these independent islands, as detailed in section B-II. This strategy significantly enhances computational efficiency without affecting the accuracy of the results.

Furthermore, an analysis of the high computation time issue of the ISRS EGN model revealed that the integral operation in the FWM efficiency factor constitutes the primary bottleneck. By decomposing the integrand, we identified three components: the high-frequency oscillation term, known as the phase mismatch factor (PMF), and two relatively stable terms, the attenuation and stimulated Raman scattering gain (SRSG). By approximating the SRSG, the remaining high-frequency oscillatory component can be directly solved, yielding a closed-form expression for the FWM efficiency factor. Further details are provided in section C-II.

### *B. The Parallel Computing Method*

In the ISRS EGN model, for a channel of interest (COI), also referred to as the channel under test (CUT), it is essential to consider the nonlinear interference from all interfering (INT) channels. In the following discussion, to avoid confusion with the frequencies $f_1$ , $f_2$ , $f$ within the channel bandwidth in Eqs. (23)-(27), the total frequencies within the WDM bandwidth are denoted as $f_1^{tot} = f_1 + k_1 R$ , $f_2^{tot} = f_2 + k_2 R$ , and $f^{tot} = f + kR$ , where $R$ is the symbol rate. When $f^{tot} = 0$ , the integration domain of the ISRS EGN model forms a diamond-shaped structure consisting of multiple islands, as shown in Fig. 2 for the seven-channel case. Here, the case where $B_{ch} > 2/3\Delta f$ is illustrated, which results in the formation of triangular islands between the diamond-shaped islands. $B_{ch}$ represents the bandwidth of a single channel, and $\Delta f$ denotes the channel spacing. The integration domain can be categorized into SCI, XCI, and MCI according to NLI types: SCI refers to interference caused by nonlinear effects within the COI itself, XCI refers to interference on the COI from one INT channel, and MCI refers to interference on the COI from two or three INT channels[31]. Notably, the EGN model provides a detailed classification of NLI, with both XCI and MCI being further subdivided into specific components. For more details, please refer to references [17].

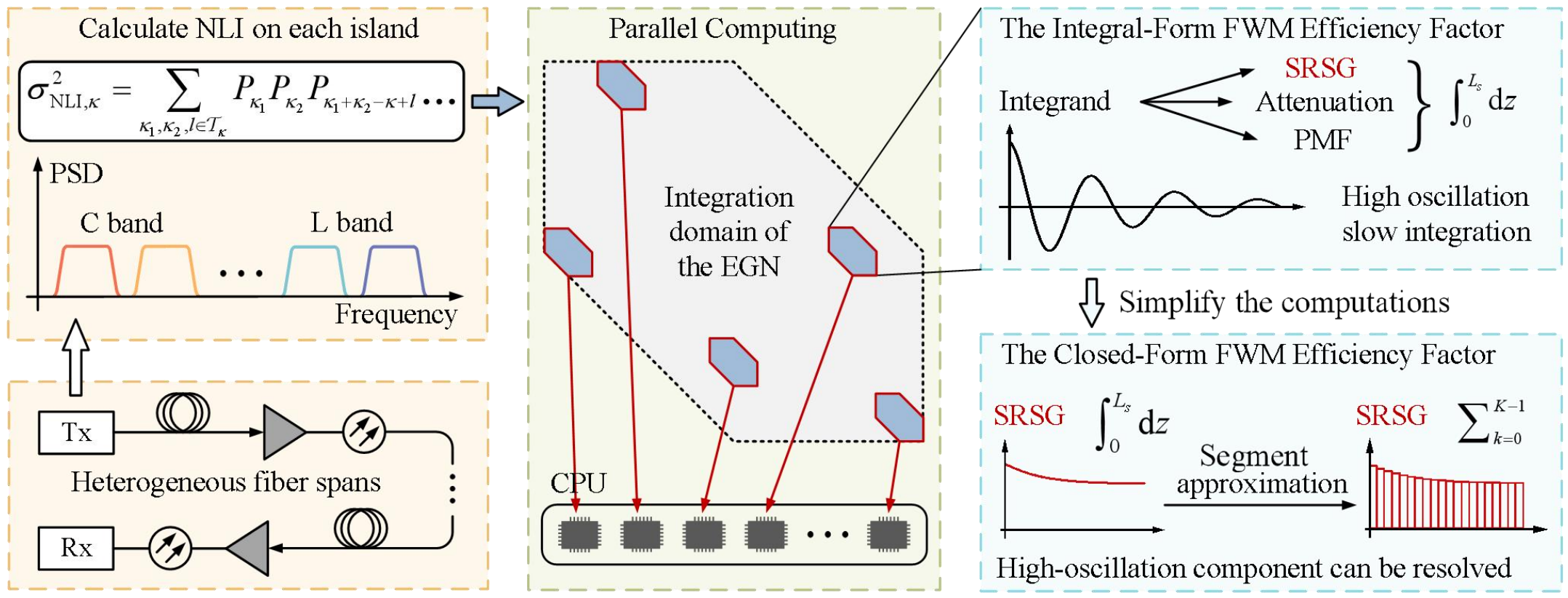

**Fig. 1.** Illustrative diagram of ISRS EGN model acceleration scheme in C+L band heterogeneous optical fiber links, including parallel computation and closed-form FWM efficiency factor.

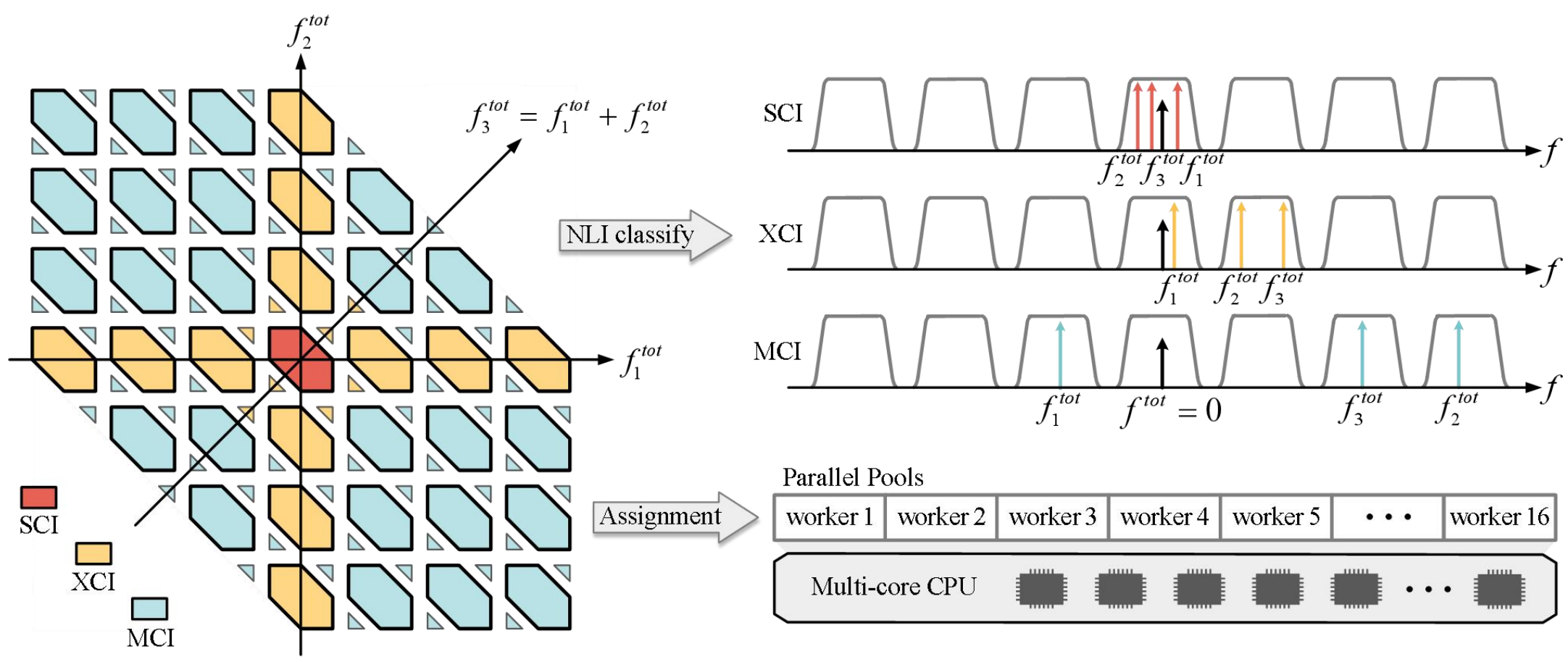

**Fig. 2.** The integration domain of the EGN model consists of numerous islands, with independent calculations on each island. Parallel computation is employed to enhance computational efficiency.

In the ISRS EGN model, integration computations are performed separately across all islands, with computations for each island being independent, making them well-suited for parallel processing. We established a parallel pool consisting of 16 workers within the program. The $P_{NLI}$ computation results for each island are stored in an array, and a summation operation is conducted once all results have been computed. It is particularly noteworthy that the inherent imbalance in computational costs among islands, caused by differences in integration multiplicity, renders traditional MATLAB `parfor` parallelization inadequate. On one hand, some workers operate for extended periods due to high-complexity computations, while others remain idle towards the end of the process [32]. On the other hand, the task-partitioning mechanism fails to effectively address load imbalance, resulting in reduced efficiency. To address these issues, this paper proposes the following optimizations: (1) An Asynchronous Dynamic Scheduling Framework, which utilizes the `parfeval` function to construct an asynchronous task queue, enabling workers to fetch new tasks in real time and allowing flexible allocation of computational resources [33]; (2) A Priority Sorting Strategy, in which tasks involving high-multiplicity integrals ( $E$ , $F$ , $G$ , and $H$ in Eq. (21)) are prioritized and placed at the front of the queue, ensuring their prompt assignment to idle workers; and (3) A Dual-Granularity Chunking Strategy, where fine-grained chunks are employed for high-multiplicity integral tasks, while coarse-grained chunks are used for regular tasks. Through a grid search, we have identified the chunk sizes that maximize performance gains. Section III-B illustrates the impact of varying chunk sizes on performance and demonstrates the effectiveness of the proposed optimization scheme.

### *C. The Closed-Form FWM Efficiency Factor*

The FWM efficiency factor quantifies the efficiency of non-degenerate FWM occurring among signals with frequencies $f_1^{tot}$ , $f_2^{tot}$ , and $f_1^{tot}+f_2^{tot}-f^{tot}$ [17]. It is calculated by dividing Eq. (1) by the effective length $L_{eff}$ . Channels with a smaller absolute value of $\left(f_1^{tot}-f^{tot}\right)\left(f_2^{tot}-f^{tot}\right)$ exhibit higher efficiency in generating FWM effects. Under a total launch power of 25 dBm, a total bandwidth of 10 THz, and 101 wavelength channels (with more detailed parameters listed in Table 1), the FWM efficiency factor across the entire integration domain following single-span transmission is calculated for the central wavelength channel, as shown in Fig. 3. Here, $f^{tot}=0$ , so the axes of maximum FWM efficiency align with the frequency axes $f_1^{tot}$ and $f_2^{tot}$. The FWM efficiency factor reaches its peak value of 1 at $f_1^{tot}=0$ and $f_2^{tot}=0$, then rapidly decays as $f_1^{tot}$ and $f_2^{tot}$ move away from $f^{tot}=0$ [31]. It should be noted that the case presented here is a simplified scenario with $C_r=0$. In the presence of ISRS, the values in the heatmap may show slight deviations, but this does not affect the subsequent analysis. Although the contribution of MCI is relatively smaller compared to SCI and XCI, it remains an essential component for accurate NLI modeling. In the context of the ISRS EGN model, neglecting the islands outside the FWM axis can introduce non-negligible errors, as demonstrated in Appendix C.

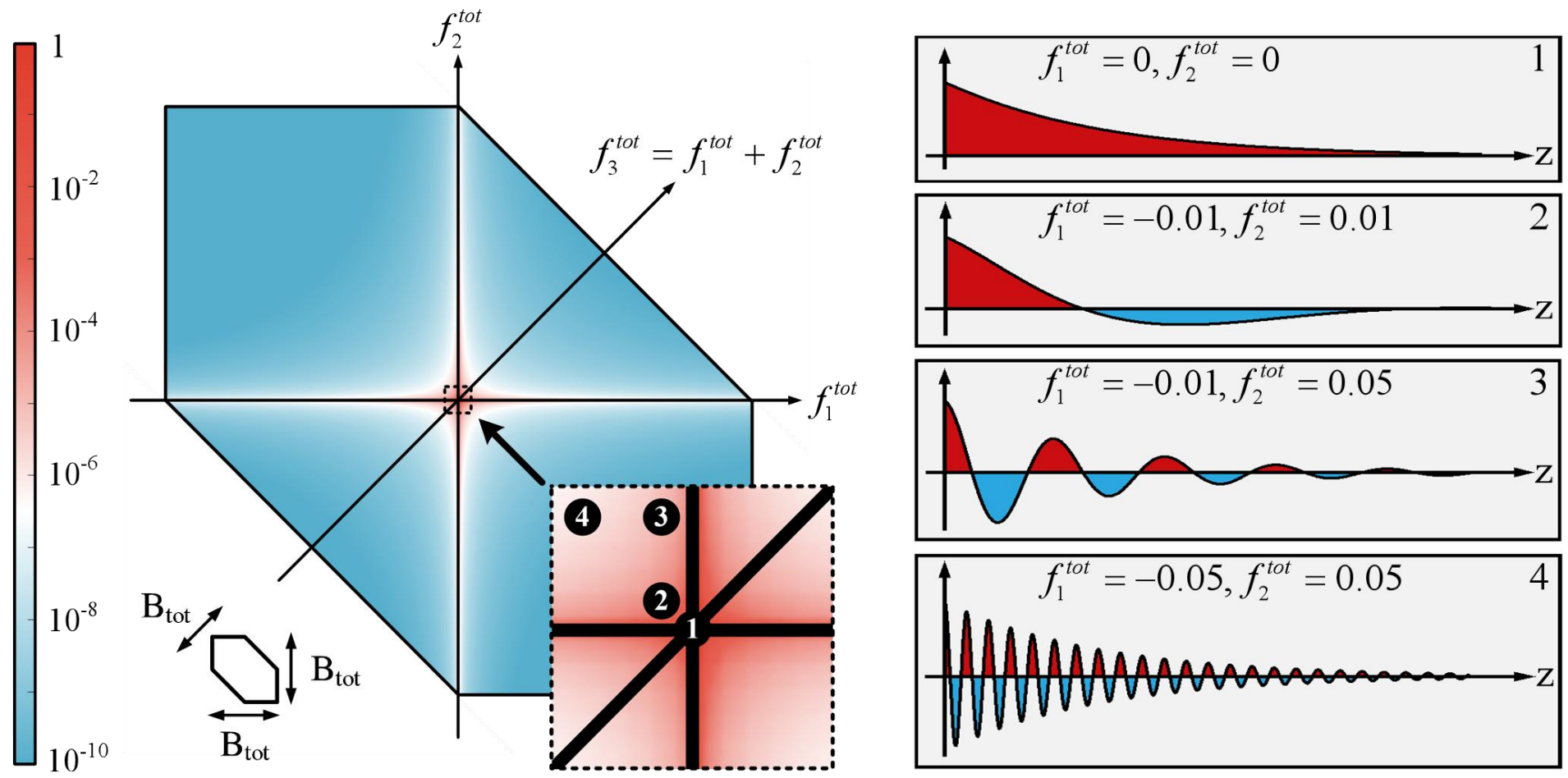

**Fig. 3.** Heatmap of the FWM efficiency factor within the EGN integration domain for a single-span WDM system with 101 wavelength channels. For four $f_1^{tot}$ and $f_2^{tot}$ (THz) positions in the figure, the waveform of the FWM efficiency factor's integrand along $z$ is shown, with the vertical axis representing amplitude and the horizontal axis representing transmission distance $z$. Positive values are depicted in red, corresponding to positive integration results, while negative values are shown in blue, indicating negative integration results.

As shown in Eq. (1), calculating the FWM efficiency factor requires integration over the span length $L_s$. In fact, for the ISRS-considered EGN model, this integration process is one of the primary factors limiting computational speed. Fig. 3 illustrates the integration process of the FWM efficiency factor at four distinct $f_1^{tot}$ and $f_2^{tot}$ positions. It can be observed that as $\left|\left(f_1^{tot}-f^{tot}\right)\left(f_2^{tot}-f^{tot}\right)\right|$ increases, the oscillation of the integrand becomes more pronounced. This higher oscillation frequency poses challenges for numerical integration, significantly increasing the computational cost required for an accurate solution. Incidentally, the high peak values and oscillatory behavior of the phased-array factor introduce another challenge, but existing solutions are already available [34].

To develop an accurate approximation method for the FWM efficiency factor, the oscillatory behavior of its integrand is initially analyzed. The FWM efficiency factor that accounts for ISRS effects, originally derived in [23], was adopted in [20], as shown in Eqs. (1)-(2). It is derived under the assumptions that the optical power is uniformly distributed across the transmission bandwidth and that the Raman gain is approximated linearly.

$$\mu_s\left(f_1,f_2,f\right)=\int_0^{L_s}\rho_s(z,f_1+f_2-f)\mathrm{e}^{i\varphi_s(f_1,f_2,f,z)}\mathrm{d}z \tag{1}$$

The integrand is decomposed into three components: the SRS gain (SRSG), the attenuation, and the phase mismatch factor (PMF), as described in Eq. (2).

$$\begin{aligned}&\rho_s(z,f_1+f_2-f)\cdot e^{i\cdot\varphi_s(f_1,f_2,f,z)}=\\&\frac{B_{tot}P_{tot}C_rL_{eff}(z)\cdot e^{-P_{tot}C_rL_{eff}(z)(f_1+f_2-f)}}{2\sinh\left(B_{tot}P_{tot}C_rL_{eff}(z)/2\right)}e^{-\alpha z}e^{i\cdot4\pi^2(f_1-f)(f_2-f)\left[\beta_{2,s}+\pi\beta_{3,s}(f_1+f_2)\right]\cdot z}\end{aligned} \tag{2}$$

where $\beta_{2,s}$ and $\beta_{3,s}$ represent the group velocity dispersion parameter and its linear slope, respectively; $B_{tot}=(2M+1)\cdot R$ is the total transmission bandwidth; $P_{tot}$ is the total launch power; and $C_r$ is the linear regression slope of the normalized Raman gain spectrum. With $f^{tot}=0$, $z=0-100$ km, and $C_r=0.028$ 1/W/km/THz, Fig. 4 illustrates the integrand of the FWM efficiency factor and its decomposed components at three different values of $\left|\left(f_1^{tot}-f^{tot}\right)\cdot\left(f_2^{tot}-f^{tot}\right)\right|$.

To facilitate explanation, only the real part of the complex result is presented in Fig. 4. The oscillatory behavior of the FWM efficiency factor integrand is attributed to the phase mismatch factor, while the SRS gain term and attenuation term remain relatively stable, forming the envelope. To derive a closed-form expression for the FWM efficiency factor, the Raman gain term was approximated, enabling the integrand to be transformed into an exponential form. The following three approximation methods are considered:

#### 1) The Maclaurin Approximation

In the derivations of various closed-form models, the Maclaurin series expansion is commonly used to simplify the integration of the FWM efficiency factor [23]. Define $\zeta=P_{tot}C_rL_{eff}$, so that Eq. (30) can be rewritten as:

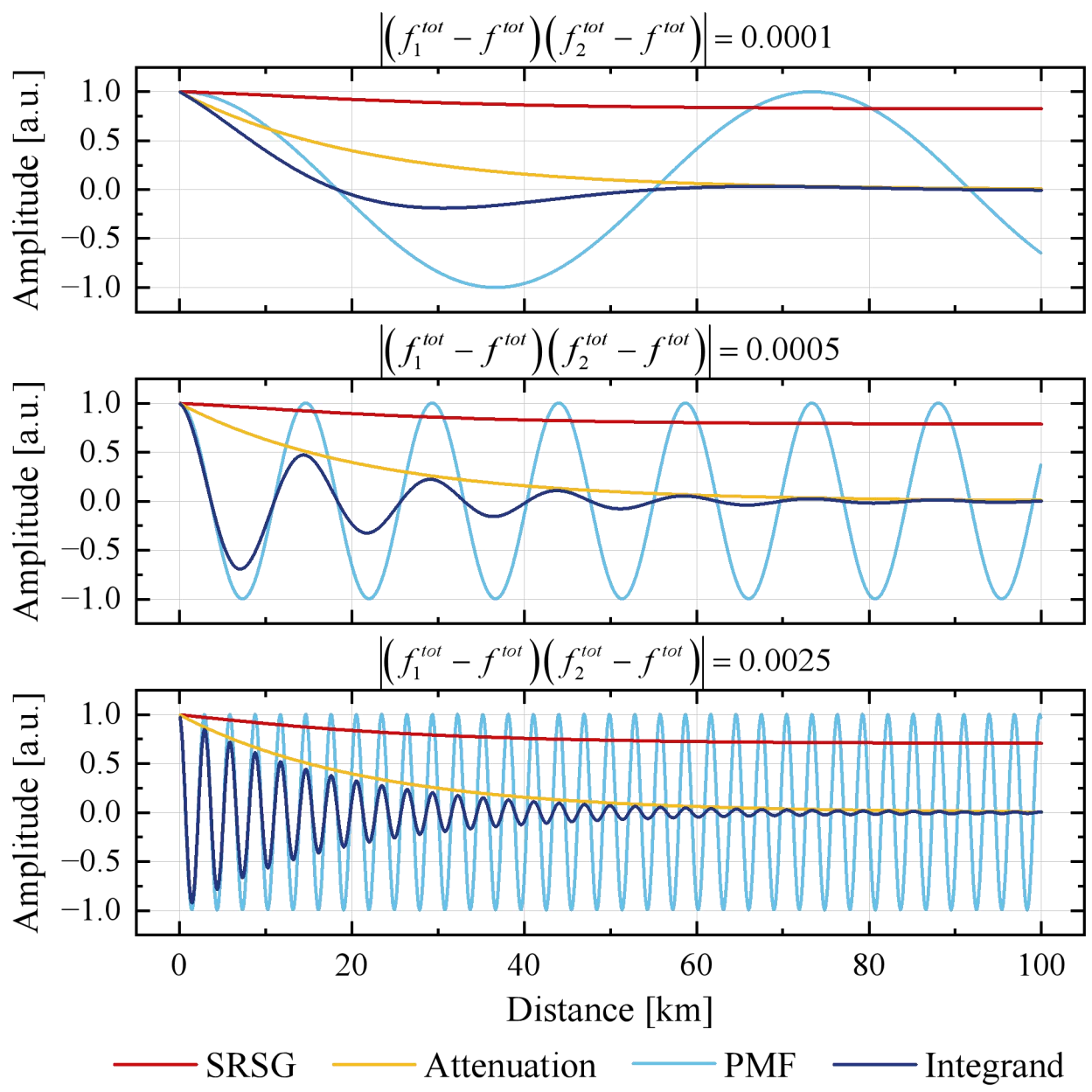

**Fig. 4.** Numerical distribution of the FWM efficiency factor integrand and its three decomposed components from $z=0$ to $z=100$ km. The overall integrand is represented by the dark blue curve.

$$\rho(z,f)=\frac{\zeta B_{tot}e^{-\zeta f}}{e^{\frac{\zeta B_{tot}}{2}}-e^{-\frac{\zeta B_{tot}}{2}}}e^{-\alpha z}, \tag{3}$$

By performing first-order and second-order Maclaurin series expansion on $e^{-\zeta f}$ and $e^{\frac{\zeta B_{tot}}{2}}$, respectively, the following can be obtained:

$$\rho(z,f)\approx(1-\zeta f)e^{-\alpha z}, \tag{4}$$

Substituting it into Eq. (1) and solving the integral yields:

$$\mu_s(f_1,f_2,f)\approx\eta-\frac{P_{tot}C_r(f_1+f_2-f)}{\alpha}\left(\eta-\frac{e^{(i\chi-2\alpha)L_s}-1}{i\chi-2\alpha}\right), \tag{5}$$

$$\chi=4\pi^2(f_1-f)(f_2-f)\left[\beta_{2,s}+\pi\beta_{3,s}(f_1+f_2)\right] \tag{6}$$

$$\eta=\frac{e^{(i\chi-\alpha)L_s}-1}{i\chi-\alpha} \tag{7}$$

In fact, the accuracy of this approximation method depends on $\mathcal{O}(P_{tot}C_rL_{eff}f)$ and $\mathcal{O}\left((P_{tot}C_rL_{eff}B_{tot}/2)^2\right)$. To a large extent, it is influenced by the $\Delta\rho(L_s)$ (see Eq. (18) for details). In C+L band scenarios, where the $B_{tot}$ typically reaches 10 THz or more, the value of $\Delta\rho(L_s)$ can become sufficiently large, leading to a loss of accuracy in the approximation. This situation arises unless the power $P_{tot}$ remains relatively low.

### 2) The Segment Approximation

To address the inaccuracies of the Maclaurin approximation for large values of $\Delta\rho(L_s)$, we propose a new approximation method. Over the integration range from 0 to $L_s$, Eq. (1) is subjected to piecewise integration into $K$ subintervals:

$$\mu_s(f_1,f_2,f)=\sum_{k=0}^{K-1}\int_{\frac{k}{K}L_s}^{\frac{k+1}{K}L_s}\rho_s(z,f_1+f_2-f)e^{i\cdot\varphi_s(f_1,f_2,f,z)}\mathrm{d}z \tag{8}$$

Since the SRS gain term exhibits low variability with respect to $z$ (as illustrated in Fig. 4), the value of $z$ within each subinterval can be approximated by its midpoint. This allows the SRS gain term to be separated from the integral and enables the integral to be directly computed. The final results are presented in Eqs. (9)-(11).

$$\mu_s(f_1,f_2,f)\approx\sum_{k=0}^{K-1}\frac{B_{tot}P_{tot}C_rL_{eff}(z_k)e^{-P_{tot}C_rL_{eff}(z_k)(f_1+f_2-f)}}{2\sinh\left(B_{tot}P_{tot}C_rL_{eff}(z_k)/2\right)}\cdot I_k \tag{9}$$

$$I_k=\frac{e^{(i\chi-\alpha)\frac{k+1}{K}L_s}-e^{(i\chi-\alpha)\frac{k}{K}L_s}}{i\chi-\alpha} \tag{10}$$

$$z_k=\frac{k+0.5}{K}L_s \tag{11}$$

The total segment count $K$ is an integer, calculated as $K=\lceil L_s/\Delta z\rceil$, where $\Delta z$ is the sampling step size. Since the SRSG remains relatively stable even for large values of $\Delta\rho(L_s)$, this approximation method demonstrates robust performance. However, its accuracy also depends on the step size. The smaller the sampling step size $\Delta z$, the greater the approximation accuracy, while the computational cost also increases correspondingly. Although the increased cost is negligible compared to integral-based methods, it remains significant when compared to the Maclaurin approximation, particularly in long-haul transmission. Furthermore, we find that the adaptive step-size strategy applied at different z positions fails to resolve this issue. This is primarily due to the relatively smooth variations in the SRS gain profile, which diminish the advantages of this strategy, particularly when considering the additional computational overhead it incurs.

### 3) The Hybrid Approximation

We believe that the most ideal approximation method should effectively combine the strengths of both the Maclaurin approximation and the Segment approximation while minimizing their respective limitations. To this end, a hybrid approximation strategy is proposed. This approach dynamically selects between the Maclaurin approximation and the Segment approximation based on the contribution of each position to the overall result, with the step size of the Segment approximation being adjusted flexibly. Specifically, as the importance of a position decreases, larger segment step sizes are utilized. When it further drops below the threshold $th$, the Maclaurin approximation method is employed to minimize computational cost, as illustrated in Fig. 5.

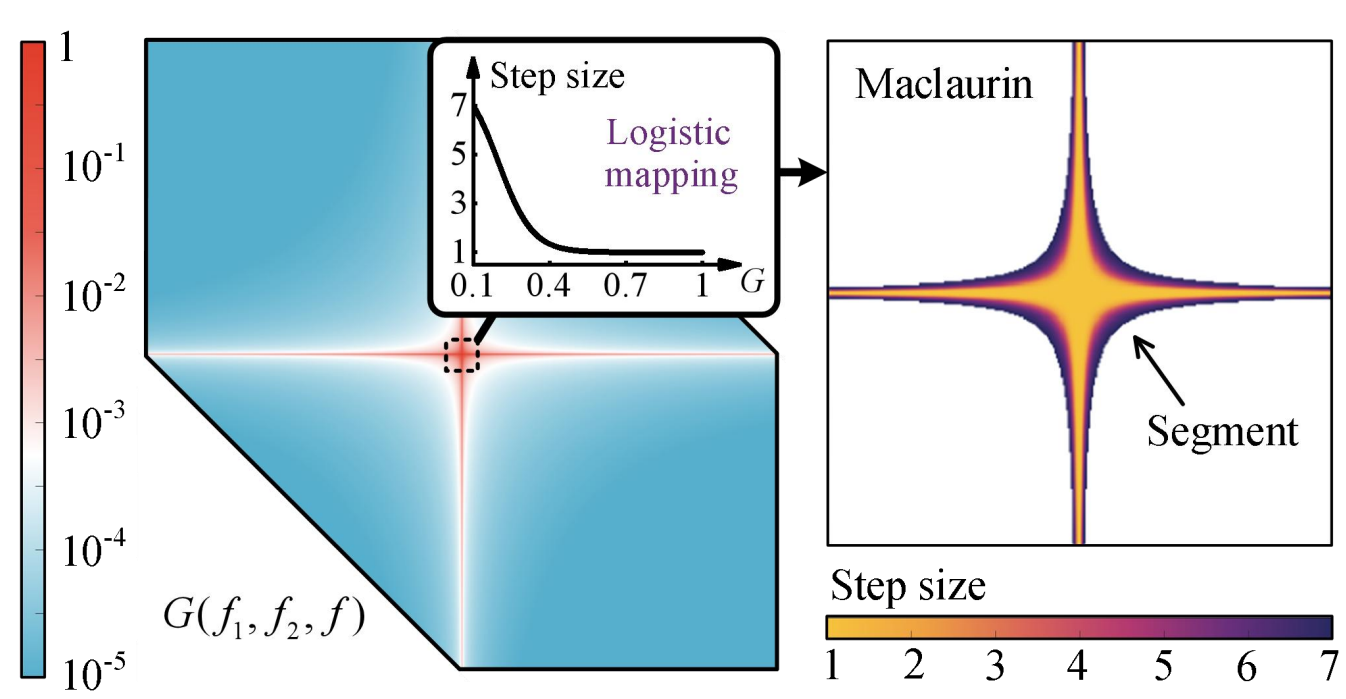

**Fig. 5.** Schematic diagram of the Hybrid approximation.

The importance function $G(f_1, f_2, f)$ is defined as:

$$G(f_1, f_2, f) = \left| \frac{1 - e^{-\alpha L_s} e^{i4\pi^2 (f_1 - f)(f_2 - f)[\beta_{2,s} + \pi\beta_{3,s}(f_1 + f_2)]L_s}}{\alpha - i4\pi^2 (f_1 - f)(f_2 - f)[\beta_{2,s} + \pi\beta_{3,s}(f_1 + f_2)]} \right| \cdot L_{eff}^{-1} \quad (12)$$

It is a special case derived from Eq. (1) under the assumption of $C_r = 0$. In fact, squaring it yields the FWM efficiency factor while neglecting the ISRS effect, as adopted in early GN models [31]. Neglecting the ISRS effect aims to obtain a normalized expression at minimal computational cost. Although this inevitably leads to some inaccuracy in importance estimation at certain positions, the impact on the final results is minimal due to the insensitivity of the step size allocation, as detailed in Appendix D.

When the value of $G(f_1, f_2, f)$ is higher than the threshold $th$, it is mapped to a specific segment step size using the logistic function $\sigma(G)$.

$$s(G) = A \cdot \sigma(G) + B \quad (13)$$

$$\sigma(G) = \frac{1}{1 + e^{k(G - G_0)}} \quad (14)$$

Here, A and B are linear correction coefficients used to accurately calibrate the mapping results to the step size range $[s_{\min}, s_{\max}]$.

$$A = \frac{s_{\max} - s_{\min}}{\sigma(th) - \sigma(1)} \quad (15)$$

$$B = s_{\max} - A \cdot \sigma(th) \quad (16)$$

A grid search was employed to identify the case with the smallest overall number of steps required, achieving a relative error in NLI coefficient of less than 0.01%. In this analysis, the number of steps in the Maclaurin approximation was set to 1. We determined the threshold $th = 0.1$, the midpoint $G_0 = 0.2$ of the logistic function, and the slope $k = 15$ .Consistently similar results were obtained when the grid search was repeated with different span counts and spectral allocations. Specifically, for the FWM efficiency factor approximation problem, this search result demonstrates strong generalization, eliminating the need to conduct a separate search for each specific scenario. For a particular type of fiber, when the span length is fixed, the value of $G(f_1, f_2, f)$ at each frequency position is fixed, which makes it more general for different link configurations and spectral assignments. Furthermore, the aforementioned parameter combinations exhibit excellent performance even under varying span lengths and fiber types, with the related results presented in Section III-D and E.

This robustness can be attributed to three key factors. First, the distribution pattern of $G$ remains relatively consistent across all cases, consistently reaching its maximum value on the FWM axis and decreasing rapidly as the $\left|(f_1^{tot} - f^{tot})(f_2^{tot} - f^{tot})\right|$ increases. Second, the importance function $G$ and the Logistic function are utilized solely to guide the selection of the approximation method and the step size, and are not directly used to compute the $\mu_s$ results. Third, even the least accurate Maclaurin approximation introduces only a limited error. The goal of the Hybrid strategy is to optimize accuracy while maintaining the speed advantage of the Maclaurin approximation. Within this framework, minimizing additional computational cost is more critical than pursuing a strictly optimal allocation strategy.

As the base model adopted in this study, one of the primary limitations of the ISRS EGN model proposed in [20] is its reliance on a linear approximation of the Raman gain spectrum. To address this issue, we employed the fitting strategy introduced in [21], which allows for the correction of the model by incorporating the actual Raman gain and attenuation spectra. Specifically, we numerically solved the Raman differential equations using the fourth-order Runge-Kutta (RK4) method to obtain the power evolution of each channel. Then, the least-squares method was used to fit the attenuation coefficients and Raman gain slopes for each wavelength channel. This process incurs an additional computation time of approximately 12 seconds. This approach overcomes several simplifying assumptions made in the derivation of the model in [20], thereby extending its applicability to a broader range of scenarios, such as non-uniform launch power distributions, wavelength-dependent attenuation, and non-triangular Raman gain spectra [21]. Notably, whether the attenuation coefficient and Raman gain slope are treated as constants or modeled with frequency dependence does not impact the acceleration method proposed in this paper, as validated in Appendix E.

Another limitation of the base model used in this study is the approximation of the nonlinear coefficient as a constant. Although this practice is common in conventional EGN models and the CFM, the resulting error is non-negligible under wide transmission bandwidths. [35] addressed this critical issue by calculating the frequency-dependent nonlinear coefficients and incorporating fitting based on experimental data. In fact, whether the frequency dependence of the nonlinear coefficients is considered or not does not affect the accelerated method proposed in this paper. Specifically, the error introduced by our acceleration approach arises solely from the approximation used in calculating the computation of the FWM efficiency, which is independent of the external nonlinear coefficient. Furthermore, although the subscripts of the relevant parameters in Eqs. (28) and (30) only include span indices, the framework inherently supports incorporating their channel dependence. During the development of the program, we designed the data structures for $\alpha$, $C_r$, and $\gamma$ in matrix form. When frequency dependence is disregarded, the corresponding elements are assigned identical values. Consequently, the numerical results related to computation time remain rigorous and are not influenced by this issue.

Moreover, although the model is a coherent accumulation model, it allows for the assignment of different optical powers to individual channels at the transmitter and supports varying amplifier gain across both channels and spans. Consequently, the model is applicable to common ISRS mitigation strategies, such as pre-compensation at the transmitter and inverse tilt compensation at each Erbium-Doped Fiber Amplifier (EDFA).

## III. Numerical Results and Discussion

### A. Parameter Settings

In this section, the two proposed ISRS EGN model acceleration methods are validated. All numerical verifications were conducted on a computer equipped with an Intel Core i7-14700KF processor (20 cores, 3.4 GHz) and 64 GB of 6400 MHz RAM. The software environment employed was MATLAB R2022b, along with the Parallel Computing Toolbox v7.7[36]. Additionally, before conducting the relevant numerical verification, we utilized the developed program to reproduce the numerical results from [20], achieving consistent outcomes. In Section III-B, we assessed the computation time of the programs executed with 1-16 workers and calculated the speedup ratio. In Section III-C, the accuracy and time-saving performance of the closed-form FWM efficiency factor were validated. Accuracy was assessed by the error in the $\eta_k$, where $k$ denotes the channel number. The calculation method detailed in Eq. (17).

$$\eta_\kappa \triangleq \frac{\sigma^2_{\mathrm{NLI},\kappa}}{P^3_{ch}} \tag{17}$$

The numerical validation in this section is conducted based on standard single-mode fiber (SMF). The launch power of each channel is set equal, with a flat spectral distribution. In the 10 THz scenario, numerical validation is carried out under two configurations, with $P_{tot}$ set to 19 dBm and 25 dBm, corresponding to $\Delta\rho(L_s)$ values of 2.0 dB and 8.2 dB, respectively, as illustrated in Fig. 6. The power transfer between outer channels, $\Delta\rho(z)$, is calculated as Eq. (18) [37].

$$\Delta\rho(z)[dB] = 4.3 \cdot P_{tot} B_{tot} C_r L_{eff}(z) \tag{18}$$

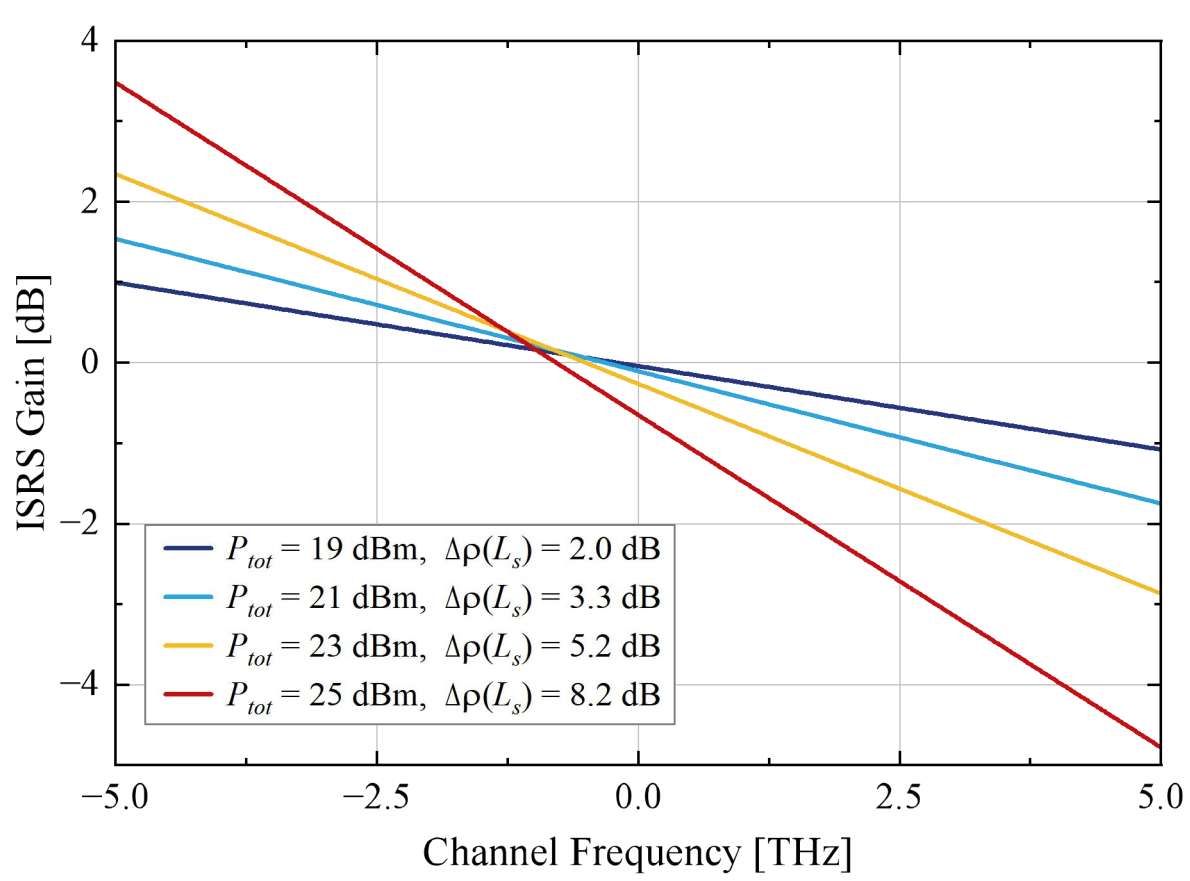

**Fig. 6.** The SRS gain versus frequency after a single 100 km span, obtained from the analytical triangular approximation with fiber attenuation compensation, based on Eq. (20) in Appendix A.

Additionally, we assessed error accumulation in long-haul transmission by validating 11 transmission distances, ranging from 1 span to 101 spans. Table 1 summarizes the parameters used in the numerical verifications. It should be noted that the numerical results presented in this paper reflect only the errors introduced by the proposed acceleration method, while the ISRS EGN model itself also exhibits inherent errors. For a detailed analysis of the error extent associated with the ISRS EGN model, please refer to [20].

TABLE I
System Parameter for Numerical Verification

| **Parameters** | **Values** |
|---|---|
| Total launch power ( $P_{tot}$ ) [dBm] | 19, 25 |
| Number of spans | 1-101, step=10 |
| Spans length ( $L_s$ ) [km] | 100 |
| Raman gain slope ( $C_r$ ) [1/W/km/THz] | 0.028 |
| Attenuation ( $\alpha$ ) [dB/km] | 0.2 |
| Dispersion ( $D$ ) [ps/nm/km] | 17 |
| Dispersion slope ( $S$ ) [ps/nm$^2$/km] | 0.067 |
| Nonlinear coefficient ( $\gamma$ ) [1/W/km] | 1.2 |
| Symbol rate ( $R$) [Gbaud] | 100 |
| Channel spacing [GHz] | 101 |
| Number of channels | 101 |

### B. Parallel Computing Performance Evaluation

Parallel computing enhances computational efficiency without affecting the calculation results, thereby significantly reducing computation time. To address the imbalance in worker resource allocation caused by uneven load distribution across islands, we use `parfeval` instead of the conventional `parfor`. It is important to note that in `parfor`-based parallel computation, MATLAB automatically determines the chunk size based on the total task volume and the number of available workers. In contrast, `parfeval` defaults to a chunk size of 1, which can result in excessive scheduling overhead and reduced computational efficiency. As shown in Fig. 7, `parfeval` with a chunk size of 1 exhibits inferior performance compared to `parfor` when executing methods with relatively low computational complexity, such as the Maclaurin approximation, the Hybrid approximation, and the Segment approximation with large step sizes (5-7km). The proposed dual-granularity chunking strategy demonstrates superior performance. The optimized chunk sizes are annotated in purple text in the far-right column of Fig. 7. For the Hybrid approximation, assigning chunk sizes of 5 and 20 to slow and fast tasks, respectively, yields optimal results, achieving a 1.34 times speedup compared to the traditional `parfor` implementation.

This work evaluated the computation time of $\eta_k$ for the central channel using 1-16 parallel workers and calculated the speedup ratios, as illustrated in Fig. 8. As the number of workers increased, the speedup ratios across all implementations of the FWM efficiency factors demonstrated a marked improvement. However, the speedup does not scale linearly with the number of workers. This behavior is consistent with Amdahl's Law, which predicts that the performance improvement of parallel computing is ultimately limited by the serial fraction of the program [38–40]. Nevertheless, with 16 workers, all methods exhibit substantial efficiency gains, achieving a speedup of approximately 11.

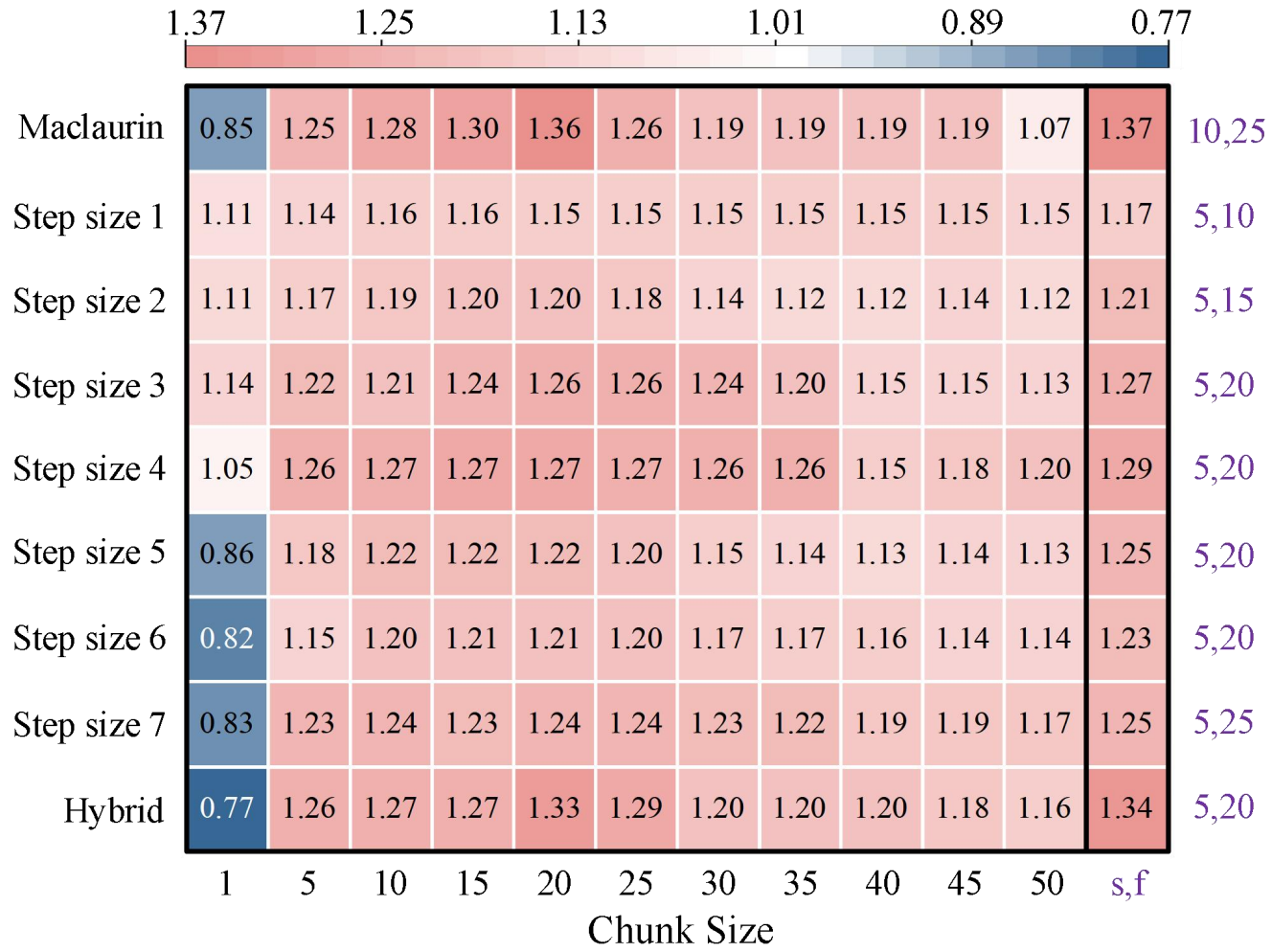

**Fig. 7.** Speedup ratios of `parfeval` with different chunk sizes relative to conventional `parfor`. The rightmost column shows the performance of the proposed dual-granularity chunking strategy, where 's' and 'f' indicate the chunk sizes for slow and fast tasks, respectively.

*C. Closed-Form FWM Efficiency Factor Evaluation*

In this section, we evaluate both the accuracy and computational efficiency of various approximation methods. The model utilizing the integral form of the FWM efficiency factor serves as the baseline for assessing the performance of the models employing closed-form approximations. Three closed-form implementations are investigated: the Segment approximation (with step sizes ranging from 1 km to 7 km), the Maclaurin approximation, and the Hybrid approximation. For each method, three modulation formats are considered: PM-QPSK, PM-16QAM, and PM-2D-Gauss.

In the scenario where $P_{tot} = 25dBm$ , the error performance is illustrated in Fig. 9. Fig. 9(a)–(c) present the error distribution versus frequency in a single-span scenario. It can be observed that the error distributions for the three modulation formats exhibit similar characteristics. Although the error associated with the Segment approximation increases with larger step sizes, its overall level remains significantly lower than that of the Maclaurin approximation. Fig. 9(d)–(f) illustrate the trend of the MAE versus the number of spans. During long-haul transmission, the MAE for all approximation methods remains relatively stable. Specifically, the MAE is approximately 0.001 dB with a step size of 1 km, increases to about 0.004 dB with a step size of 2 km, while the Hybrid approximation lies between these values at approximately 0.0033 dB. These results demonstrate that even under high $\Delta\rho(L_s)$ scenarios, both the Segment approximation with smaller step sizes and the Hybrid approximation deliver highly accurate results, with negligible errors.

When $P_{tot}$ decreases to 19 dBm, the error performance is illustrated in Fig. 10. Compared to the scenario at $P_{tot} = 25dBm$ , the error of the Maclaurin approximation is significantly reduced, thereby narrowing the gap with the Segment approximation. During long-haul transmission, the MAE of the Maclaurin approximation remains below 0.012 dB. Although the error of the Segment approximation is also reduced relative to its performance at $P_{tot} = 25dBm$ , its precision advantage over the Maclaurin approximation is notably diminished. This is particularly evident when the segment step size is set to 7 km, where its MAE exceeds that of the Maclaurin approximation. Additionally, the Hybrid approximation consistently achieves an MAE between those of the Segment approximation with step sizes of 1 km and 2 km, maintaining a value of approximately 0.0007 dB.

Next, the time-saving performance of the different approximation methods is evaluated. As shown in Fig. 11(a), the Segment approximation provides a substantial speedup compared to the full integration model. In the single-span scenario, the Segment approximation with a step size of 1 km achieves a 98.1% reduction in computation time, which further improves to 98.5% in the case of 101 spans. As the step size increases, the computation time decreases even further. Fig. 11(b) indicates that this effect becomes more pronounced for long-haul transmissions. In fact, the computation time of all approximation methods increases approximately linearly with the number of spans increases, but their growth rates differ significantly. For the Segment approximation, the computation time increases at a rate of 187.5 seconds per span when using a step size of 1 km, which decreases to 33.3 seconds per span when a step size of 7 km. Despite the substantial improvements in efficiency, this comes at the cost of reduced accuracy. Moreover, its computation time remains considerably higher than that of the Maclaurin approximation, which achieves a much lower growth rate of just 3.3 seconds per span.

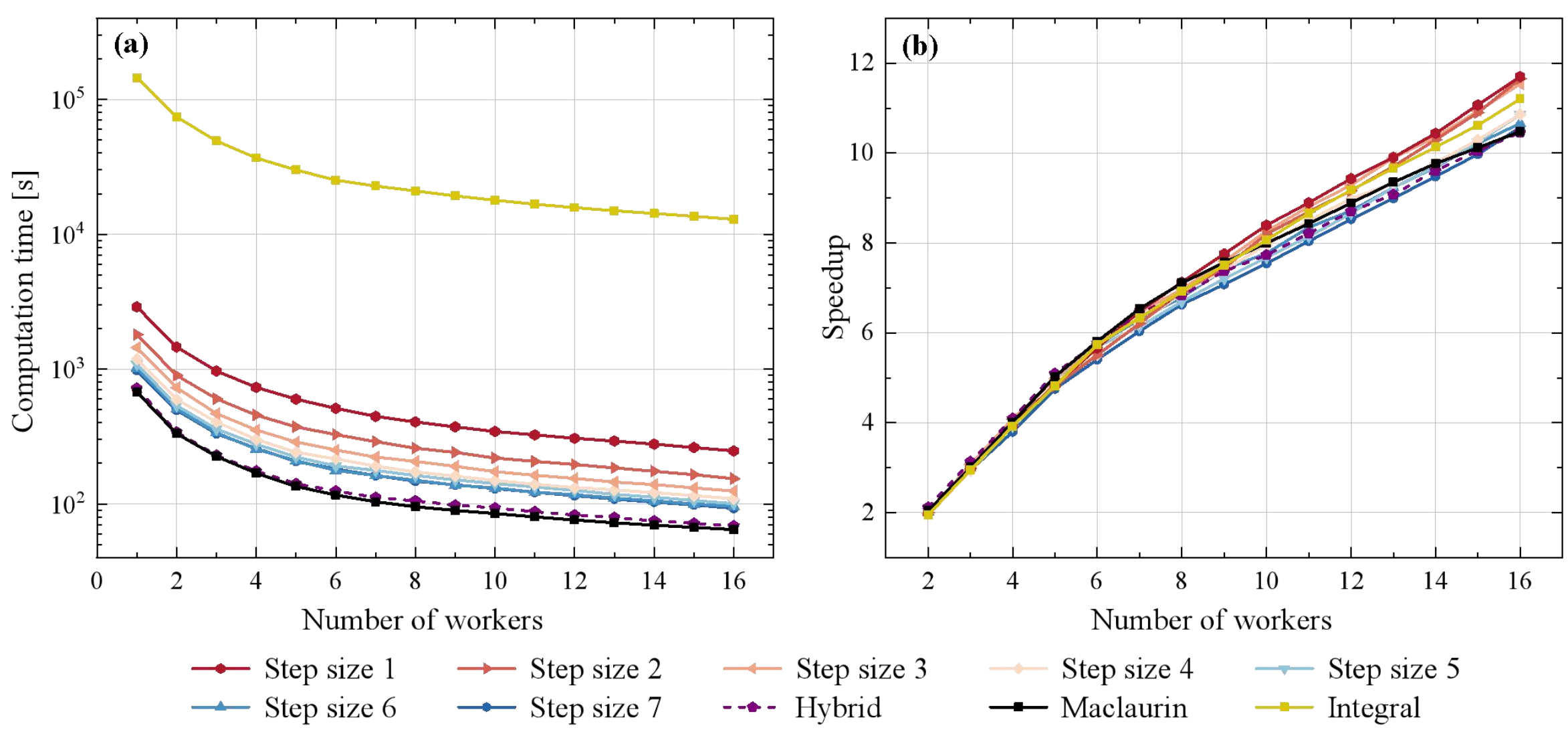

**Fig. 8.** (a) Computation time of the central wavelength channel using the full integral model and three approximation methods, under configurations ranging from 1 to 16 workers. The Segment approximation is evaluated with step sizes varying from 1 km to 7 km. (b) Speedup performance with multiple workers.

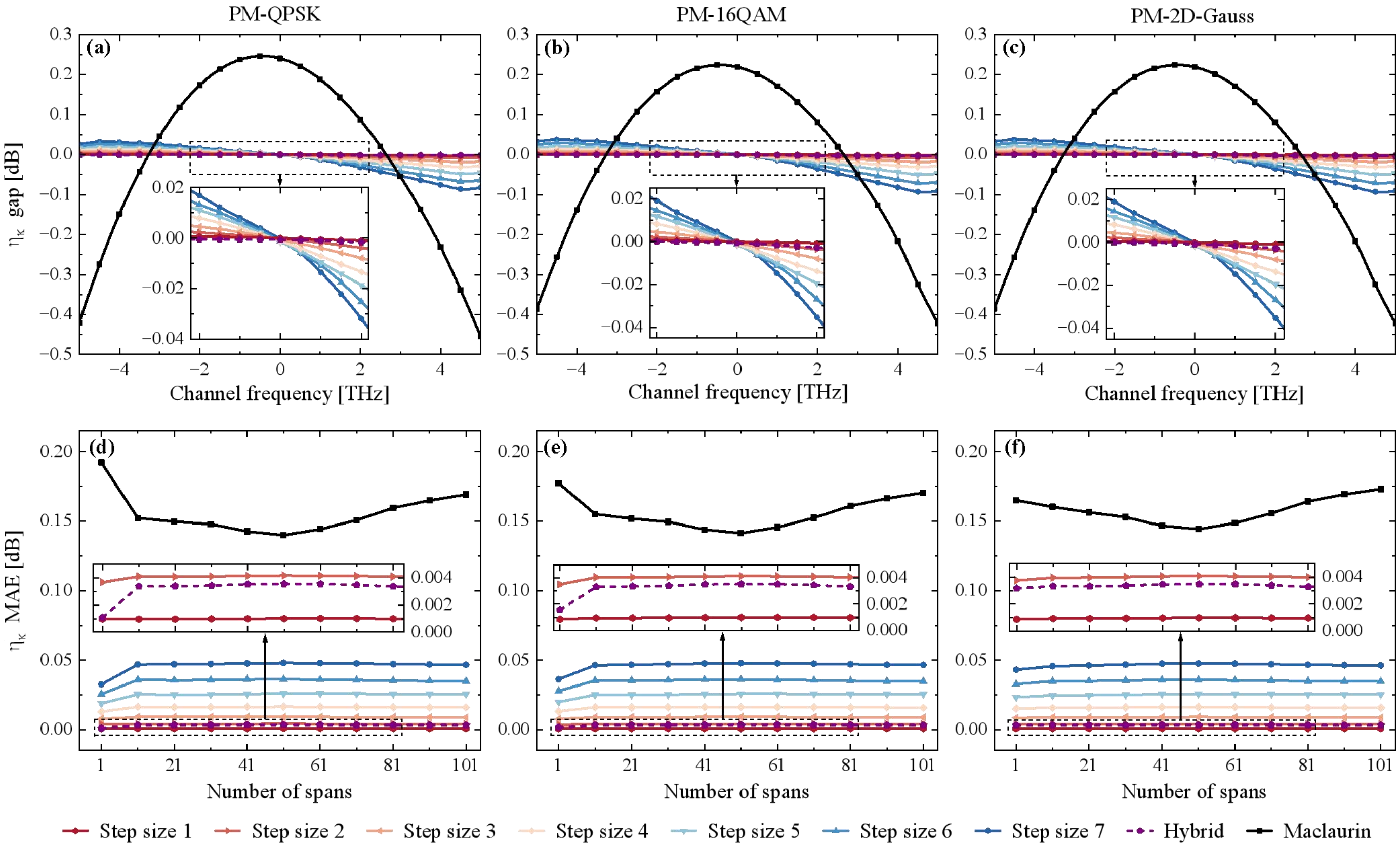

**Fig. 9.** Errors of various approximation methods relative to the integral result when $P_{tot} = 25dBm$ .

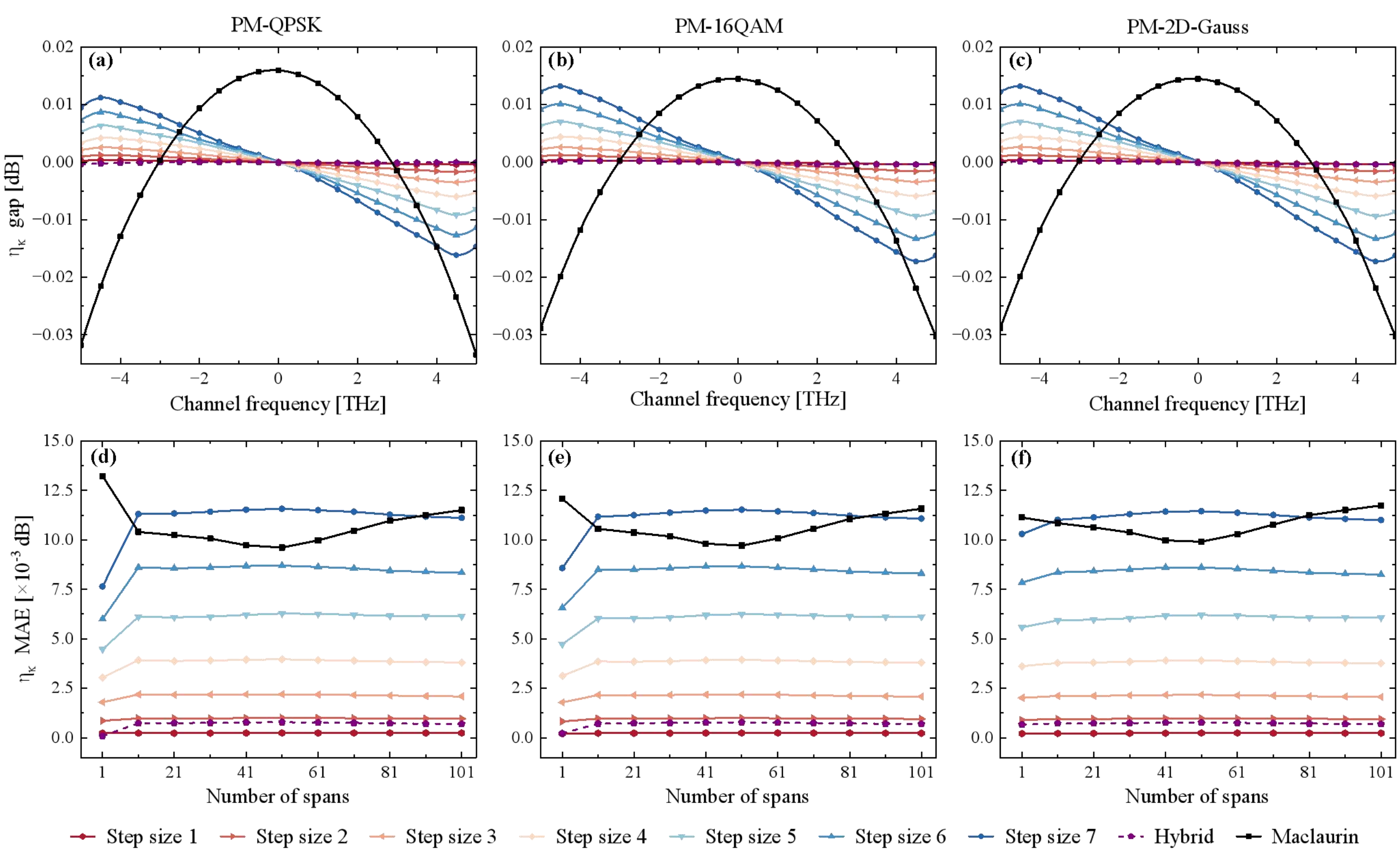

**Fig. 10.** Errors of various approximation methods relative to the integral result when $P_{tot} = 19dBm$ .

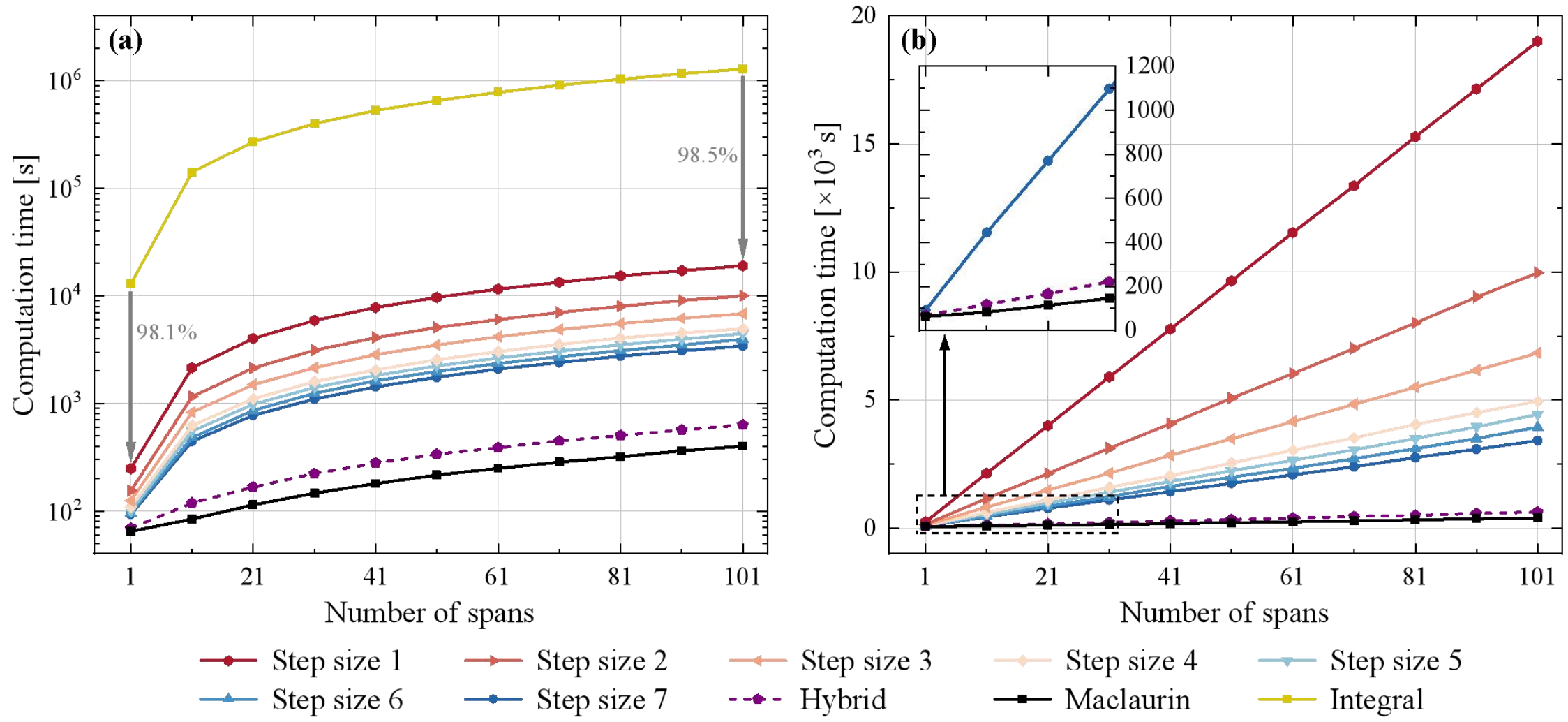

**Fig. 11.** Computation time of various approximation methods for the central channel at different transmission distances. (a) Results on a logarithmic scale, including additional data showing the computation time of the full integration model for comparison. (b) Results on a linear scale, highlighting the relative performance differences among the approximation methods.

In contrast to the pure Segment approximation with a fixed step size, the Hybrid approximation demonstrates a notable advantage. While maintaining high accuracy, its computation time increases at a significantly lower rate of just 5.6 seconds per span. Particularly, in single-span scenarios, the hybrid approximation requires only 6.5% more computation time than the Maclaurin approximation, while achieving a 99.5% reduction in computation time compared to the full-integration model. In summary, the Hybrid approximation effectively overcomes the accuracy limitations of the Maclaurin approximation while incurring only a minor increase in computational cost. This makes it a precise and efficient solution, particularly well-suited for scenarios where the impact of ISRS is significant.

## *D. Evaluation with Varying Span Lengths*

When the modulation format is QPSK, the MAE of all channels after single-span transmission at various span lengths is illustrated in Fig. 12. Across different span lengths, the Hybrid approximation consistently demonstrates very low errors, closely aligning with the performance of the Segment approximation with a step size of 1 km.

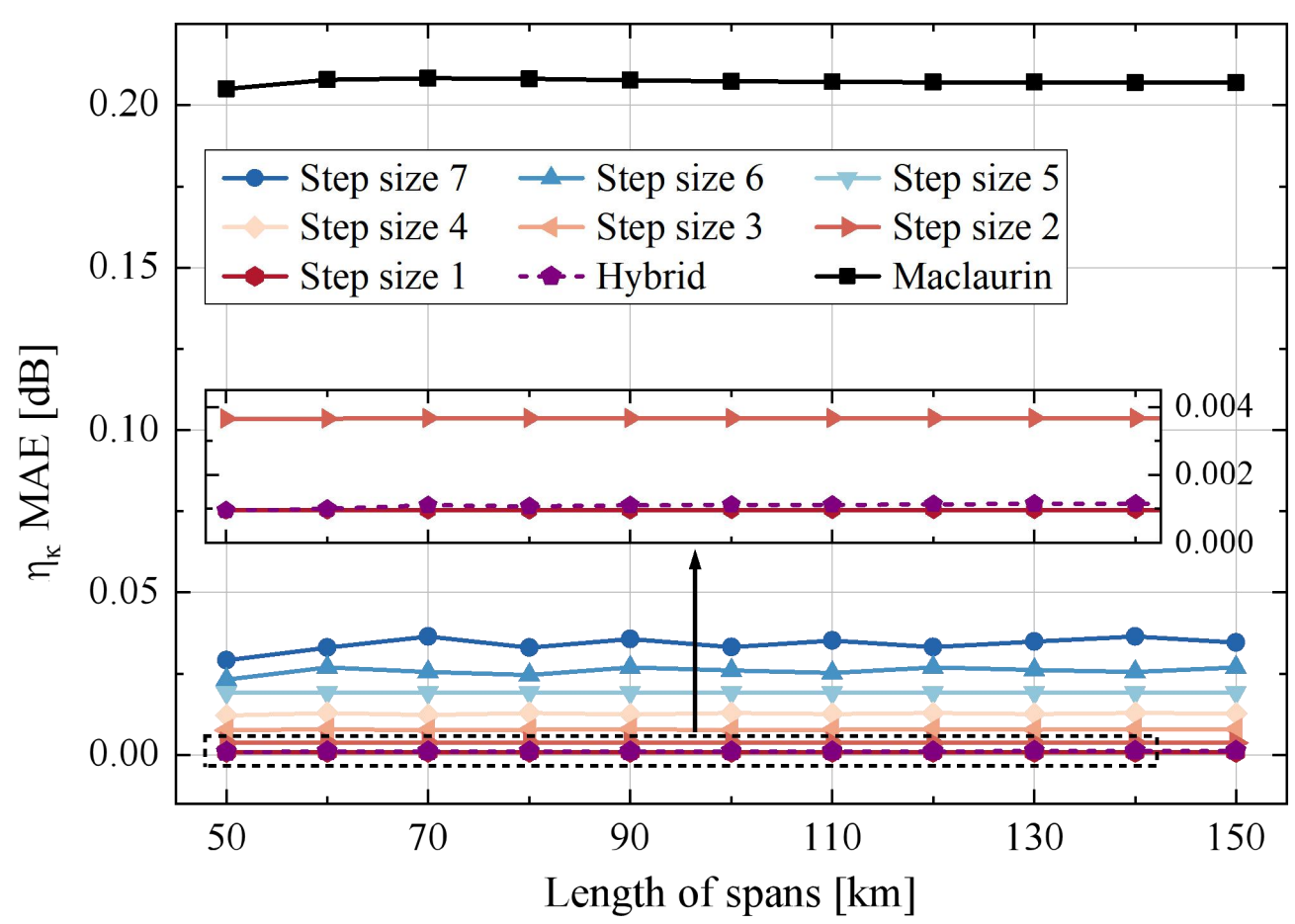

**Fig. 12.** Errors of various approximation methods across span lengths of 50 to 150 km.

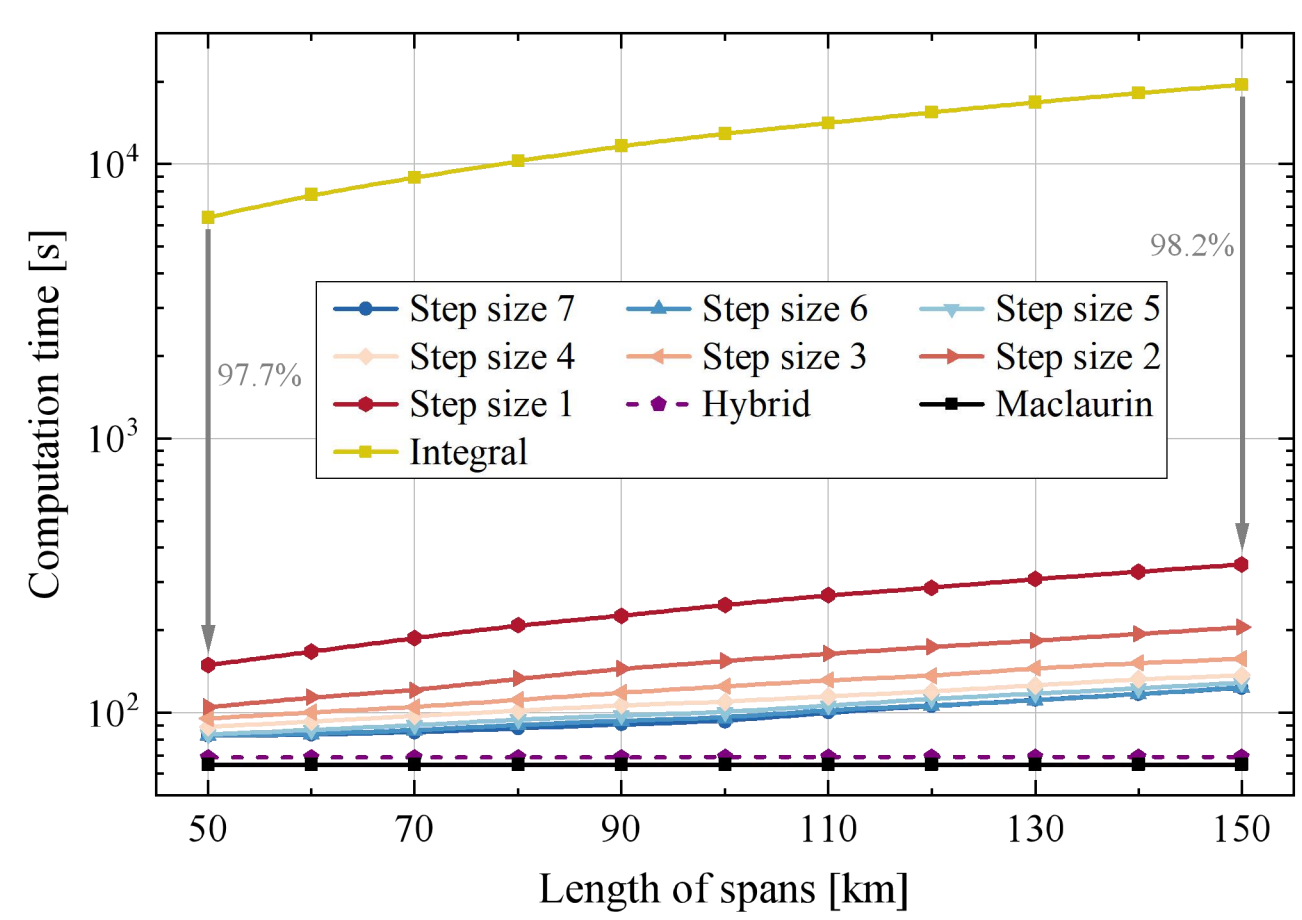

**Fig. 13.** Computation time of various approximation methods across span lengths of 50 to 150 km.

As illustrated in Fig. 13, the computation time for both the integral method and the Segment approximation increases with span length. In contrast, the Maclaurin approximation remains unaffected, and the impact on the Hybrid approximation is nearly negligible. Across all span lengths, the Hybrid approximation consistently achieves a computation speed that is very close to that of the Maclaurin approximation.

### *E. Evaluation with Different Fiber Types*

The performance of various approximation methods was further evaluated on two additional types of optical fibers, with relevant parameters obtained from [9]. As shown in Fig. 14, the proposed Hybrid approximation consistently demonstrates very low errors across all three fiber types. Notably, for PSCF, the Hybrid approximation achieves a lower error than the Segment approximation with a step size of 1 km for the first time, benefiting from the cancellation of positive and negative errors between the Segment and Maclaurin approximations. Since the type of optical fiber does not influence computational speed, the corresponding runtime results are not presented.

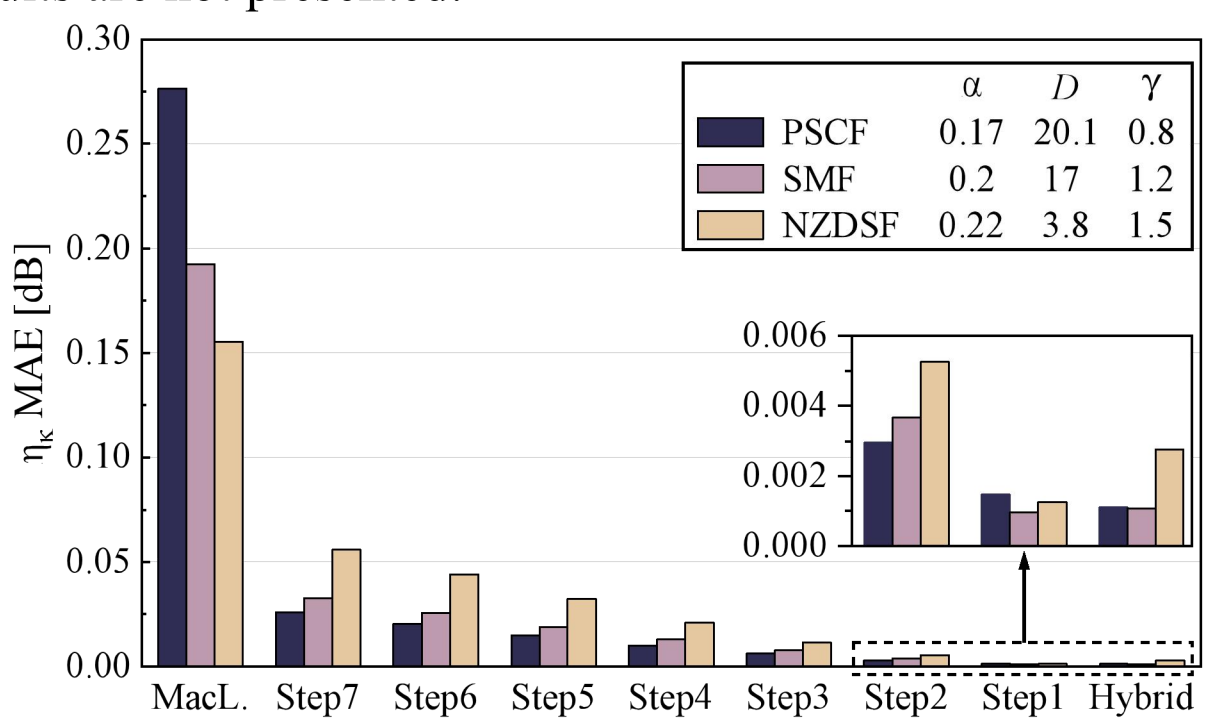

**Fig. 14.** Errors of various approximation methods for three types of optical fibers. The relevant parameters are annotated in the figure, where units of $\alpha$, $D$, and $\gamma$ are [dB/km], [ps/nm/km], and [1/W/km], respectively.

## IV. Conclusion

This work was conducted to address the issue of the slow computation speed of the ISRS EGN model. We identified that the primary factor contributing to the long computation time in the ISRS EGN model is the integration operations required for calculating the FWM efficiency factor. By analyzing the oscillatory behavior of the integrand, we proposed an accurate approximation method and derived a closed-form expression of the FWM efficiency factor. This approach demonstrates exceptional performance, maintaining the MAE below 0.0036 dB while simultaneously achieving high computational speed. In addition, we proposed a parallel computing method and performed targeted optimizations, which, based on the proposed approximation method, further improve computational speed and result in significant time savings.

The findings of this work have substantial practical implications. When existing closed-form expressions fail to meet accuracy requirements, the proposed method enables high-precision calculations of the ISRS EGN model with relatively low computational cost. In addition, this work introduces a more efficient data generation approach for machine learning applications, facilitating the creation of larger datasets within the same time frame. Future research will extend this work to further enhance the computational speed of the ISRS EGN model.

## Appendix

### *A. ISRS*

Stimulated Raman scattering is a common nonlinear effects in optical fibers, causing a portion of the incident light's power is transferred from the original beam to another beam with a lower frequency[41-42]. In multi-channel transmission systems, the frequency differences between channels give rise to the ISRS effect[43-44]. Assuming attenuation variations are negligible, the normalized signal power distribution of spectral component $f$ can be solved as:

$$\rho(z,f)=\frac{B_{tot}P_{tot}C_rL_{eff}(z)\cdot e^{-\alpha z-P_{tot}C_rL_{eff}(z)f}}{2\sinh\left(P_{tot}B_{tot}C_rL_{eff}(z)/2\right)},\tag{19}$$

by excluding fiber attenuation from Eq. (19), the SRS gain can be obtained as:

$$SRS_G(z,f)=\frac{B_{tot}P_{tot}C_rL_{eff}(z)\cdot e^{-P_{tot}C_rL_{eff}(z)f}}{2\sinh\left(P_{tot}B_{tot}C_rL_{eff}(z)/2\right)}\tag{20}$$

### *B. ISRS EGN Model*

The main equation of the ISRS EGN model proposed in [20] is given in Eq. (21):

$$\begin{aligned}\sigma^2_{\mathrm{NLI},\kappa}=&\sum_{\kappa_1,\kappa_2,l\in\mathcal{T}_\kappa}P_{\kappa_1}P_{\kappa_2}P_{\kappa_1+\kappa_2-\kappa+l}\big(D_\kappa+\Phi_{\kappa_1}\cdot\delta_{\kappa_1,\kappa_1+\kappa_2-\kappa+l}E_\kappa\\&+\Phi_{\kappa_2}\delta_{\kappa_2,\kappa_2+\kappa_1-\kappa+l}F_\kappa+\Phi_{\kappa_1}\delta_{\kappa_1,\kappa_2}G_\kappa+\delta_{\kappa_1,\kappa_2}\delta_{\kappa_2,\kappa_2+\kappa_1-\kappa+l}\Psi_{\kappa_1}H_\kappa\big),\end{aligned}\tag{21}$$

Within the integration domain, each island is delineated by $\kappa_1$, $\kappa_2$, and $l$, with their combinations recorded in $\mathcal{T}_\kappa$:

$$\mathcal{T}_\kappa=\{(\kappa_1,\kappa_2,l)\in\{-M,\ldots,M\}^2\times\{-1,0,1\}:-M\le\kappa_1+\kappa_2-\kappa+l\le M\},\tag{22}$$

The $D_\kappa$ in Eq. (21) can be used to obtain the results for the ISRS GN model, and the correction results are calculated by $E_\kappa$, $F_\kappa$, $G_\kappa$, and $H_\kappa$. The specific calculation method is provided in Eqs. (23)-(27):

$$\begin{aligned}D_\kappa(\kappa_1,\kappa_2,l)=&\frac{16}{27}R^3\int_{-R/2}^{R/2}\mathrm{d}f\int_{-R/2}^{R/2}|S(f_1)|^2\,\mathrm{d}f_1\\&\cdot\int_{-R/2}^{R/2}|S(f_2)|^2|S(f_1+f_2-f-lR)|^2|\Upsilon(f_1+\kappa_1R,f_2+\kappa_2R,f+\kappa R)|^2\,\mathrm{d}f_2\end{aligned}\tag{23}$$

$$\begin{aligned}E_\kappa(\kappa_1,\kappa_2,l)=&\frac{16}{27}R^2\int_{-R/2}^{R/2}\mathrm{d}f\int_{-R/2}^{R/2}S(f_1)\mathrm{d}f_1\\&\cdot\int_{-R/2}^{R/2}|S(f_2)|^2\,S^*(f_1+f_2-f-lR)\Upsilon(f_1+\kappa_1R,f_2+\kappa_2R,f+\kappa R)\mathrm{d}f_2\\&\cdot\int_{-R/2}^{R/2}S^*(f_1')S(f_1'+f_2-f-lR)\Upsilon^*(f_1'+\kappa_1R,f_2+\kappa_2R,f+\kappa R)\mathrm{d}f_1'\end{aligned}\tag{24}$$

$$\begin{aligned}F_\kappa(\kappa_1,\kappa_2,l)=&\frac{32}{81}R^2\int_{-R/2}^{R/2}\mathrm{d}f\int_{-R/2}^{R/2}|S(f_1)|^2\,\mathrm{d}f_1\\&\cdot\int_{-R/2}^{R/2}S(f_2)S^*(f_1+f_2-f-lR)\Upsilon(f_1+\kappa_1R,f_2+\kappa_2R,f+\kappa R)\mathrm{d}f_2\\&\cdot\int_{-R/2}^{R/2}S^*(f_2')S(f_1+f_2'-f-lR)\Upsilon^*(f_1+\kappa_1R,f_2'+\kappa_2R,f+\kappa R)\mathrm{d}f_2'\end{aligned}\tag{25}$$

$$\begin{aligned}G_\kappa(\kappa_1,\kappa_2,l)=&\frac{16}{81}R^2\int_{-R/2}^{R/2}\mathrm{d}f\int_{-R/2}^{R/2}S(f_1)\mathrm{d}f_1\\&\cdot\int_{-R/2}^{R/2}S(f_2)|S(f_1+f_2-f-lR)|^2\,\Upsilon(f_1+\kappa_1R,f_2+\kappa_2R,f+\kappa R)\mathrm{d}f_2\\&\cdot\int_{-R/2}^{R/2}S^*(f_1')S^*(f_1+f_2-f_1')\Upsilon^*(f_1'+\kappa_1R,f_1+f_2-f_1'+\kappa_2R,f+\kappa R)\mathrm{d}f_1'\end{aligned}\tag{26}$$

$$
\begin{aligned}
H_\kappa(\kappa_1,\kappa_2,l) &= \frac{16}{81}R\int_{-R/2}^{R/2}\mathrm{d}f\int_{-R/2}^{R/2}S(f_1)\mathrm{d}f_1\int_{-R/2}^{R/2}S^*(f_1')\mathrm{d}f_1' \\
&\cdot\int_{-R/2}^{R/2}S(f_2)S^*(f_1+f_2-f-lR)\Upsilon(f_1+\kappa_1R,f_2+\kappa_2R,f+\kappa R)\mathrm{d}f_2 \\
&\cdot\int_{-R/2}^{R/2}S^*(f_2')S(f_1'+f_2'-f-lR)\Upsilon^*(f_1'+\kappa_1R,f_2'+\kappa_2R,f+\kappa R)\mathrm{d}f_2'
\end{aligned}
\quad (27)
$$

In Eqs. (23)-(27), the calculation method for the link function $\Upsilon$ is as follows:

$$
\begin{aligned}
\Upsilon(f_1,f_2,f) &= \sum_{s=1}^{N_{sp}}\gamma_s\mu_s(f_1,f_2,f)\cdot \mathrm{e}^{\imath 4\pi^2(f_1-f)(f_2-f)\sum_{s'=1}^{s-1}(\beta_{2,s'}L_{s'}+\pi(f_1+f_2)\beta_{3,s'}L_{s'})} \\
&\cdot\prod_{s'=1}^{s-1}g_{s'}^{3/2}\sqrt{\rho_{s'}(L_{s'},f_1)}\sqrt{\rho_{s'}(L_{s'},f_1-f+f_2)}\sqrt{\rho_{s'}(L_{s'},f_2)}\cdot\prod_{s'=s}^{N_{sp}}g_{s'}^{1/2}\sqrt{\rho_{s'}(L_{s'},f)}
\end{aligned}
\quad (28)
$$

Here, $\gamma_s$ represents the nonlinear coefficient, and $\mu_s(f_1,f_2,f)$ reflects the efficiency of FWM. Since $\rho_s$ accounts for the impact of ISRS, $\mu_s$ is in integral form, as shown in Eq. (1). Where:

$$\varphi_s(f_1,f_2,f,z) = 4\pi^2(f_1-f)(f_2-f)\left[\beta_{2,s}+\pi\beta_{3,s}(f_1+f_2)\right]z \quad (29)$$

$$\rho_s(z,f) = \frac{B_{\mathrm{tot}}P_{\mathrm{tot}}C_rL_{\mathrm{eff}}(z)\mathrm{e}^{-\alpha_s z-P_{\mathrm{tot}}C_rL_{\mathrm{eff}}(z)f}}{2\sinh\left(P_{\mathrm{tot}}B_{\mathrm{tot}}C_rL_{\mathrm{eff}}(z)/2\right)} \quad (30)$$

### C. *Evaluation of Removing Off-Axis FWM Islands*

Results for three modulation formats after single-span transmission are presented in Fig. 15. Ignoring off-axis FWM islands significantly underestimates NLI, introducing non-negligible errors.

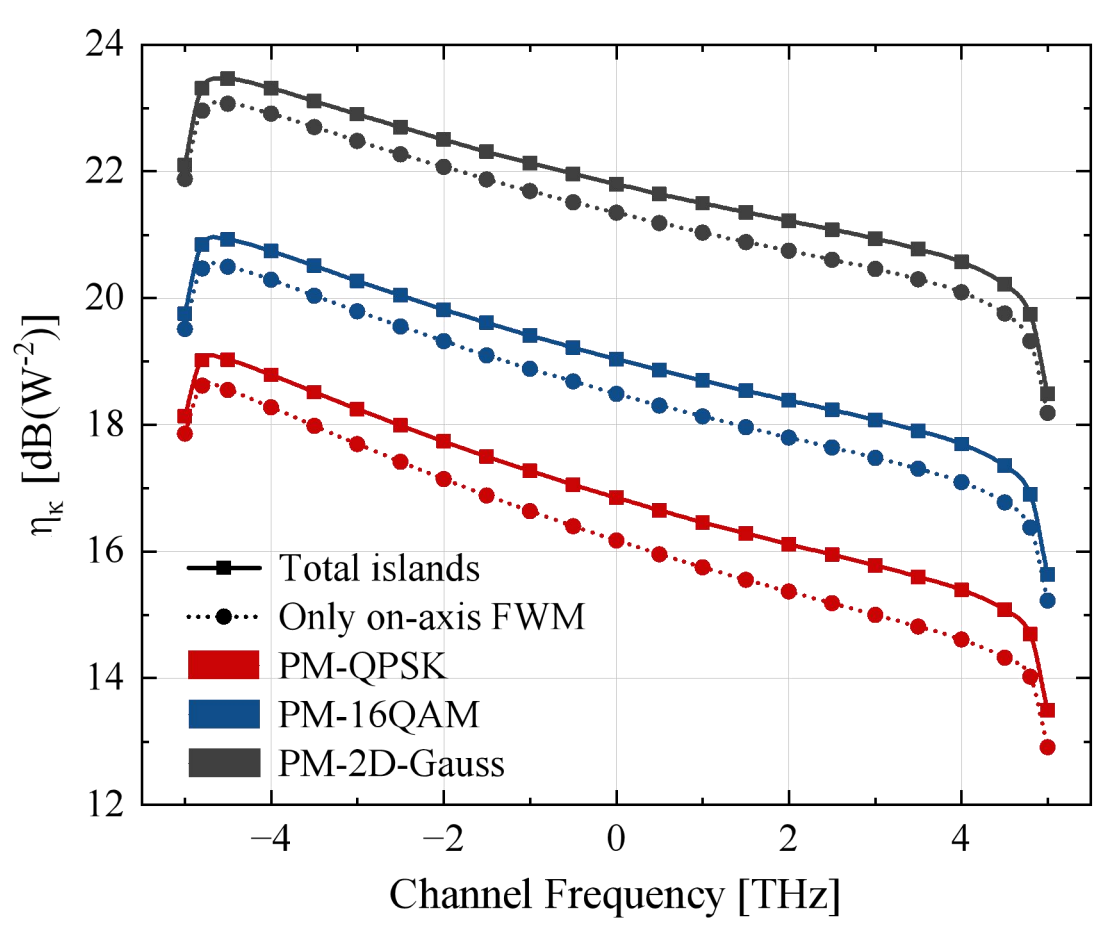

**Fig. 15.** Comparison of calculation results between the full integration domain and the on-axis FWM islands only.

### D. *Neglecting ISRS Effects in the Importance Function*

For the Hybrid approximation, the optimal strategy is to assign sufficiently fine segment step sizes to all regions with high FWM contributions. These critical regions occupy only a very small portion of the overall integration domain. Consequently, the overall computational speed remains virtually unaffected, while the approximation accuracy improves significantly. In fact, this strategy largely results in the insensitivity to the specific allocation of step sizes. Specifically, the important region along the FWM axis covering 5% of the total transmission bandwidth is considered. As shown in Fig. 16, significant differences between $G$ and $|\mu_s|/L_{eff}$ appear only near the crossing positions of the FWM axis. However, at these positions, the FWM efficiency remains high regardless of whether ISRS effects are considered, which means the segment step size should be set to 1. Moreover, the inherent normalization property of $G$ provides substantial advantages in terms of algorithmic generality and robustness. In contrast, when $\mu_s$ serves as the importance function, normalization not only adds extra computational cost but also largely removes the ISRS effects that $\mu_s$ aims to represent.

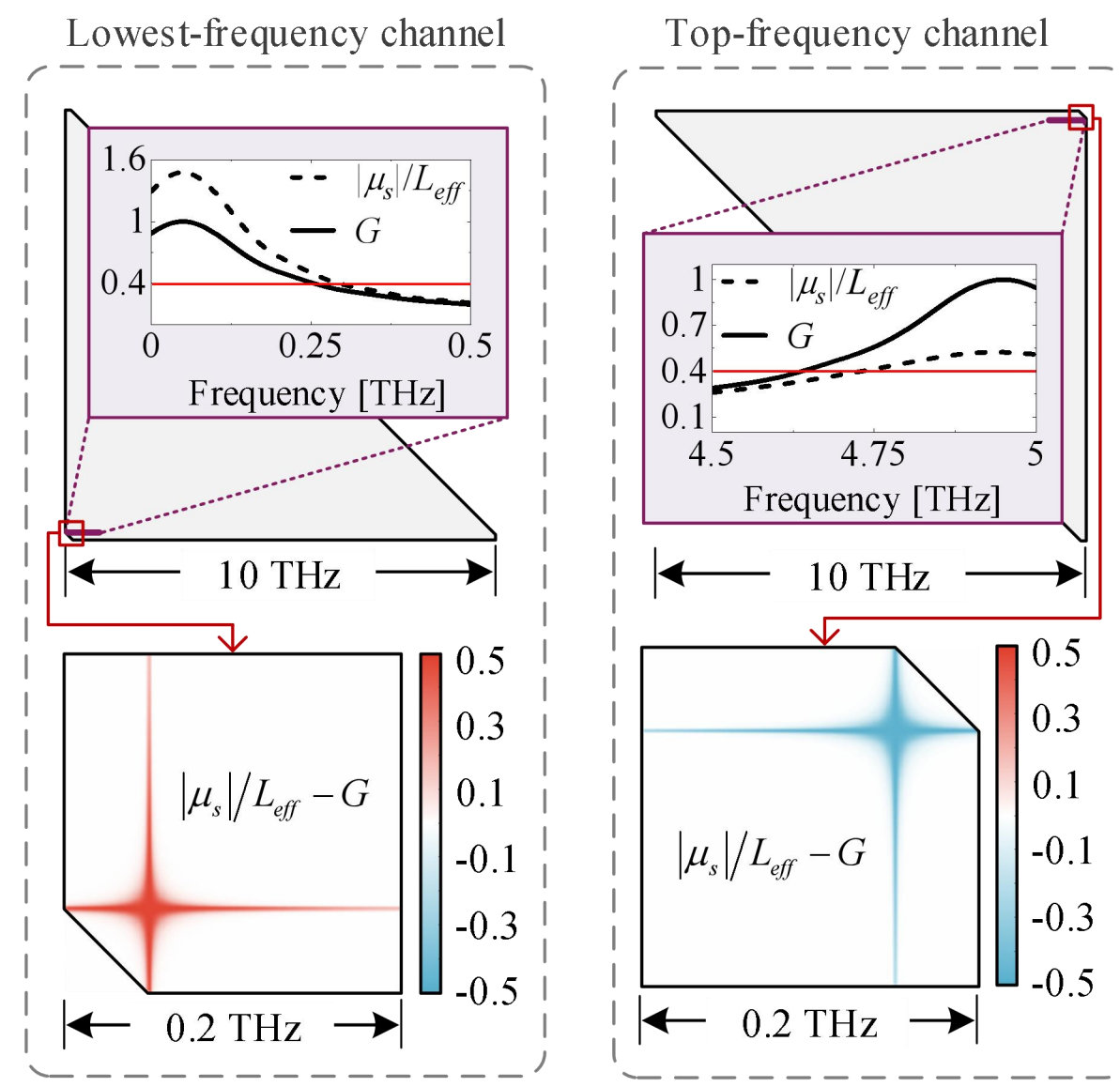

**Fig. 16.** Comparative analysis of importance functions constructed using $\mu_s$ and $G$.

### E. *Evaluation with Actual Attenuation and Raman Gain Spectra*

When the actual attenuation and Raman gain spectra are used, the error performance of the approximation methods does not change significantly. Fig. 17 presents the case with $P_{tot}=25dBm$ and PM-QPSK modulation format, which shows results very similar to Fig. 9(d). Moreover, the use of actual attenuation and Raman gain spectra does not directly affect the computational speed of the approximation methods, and therefore the computation times are not reported.

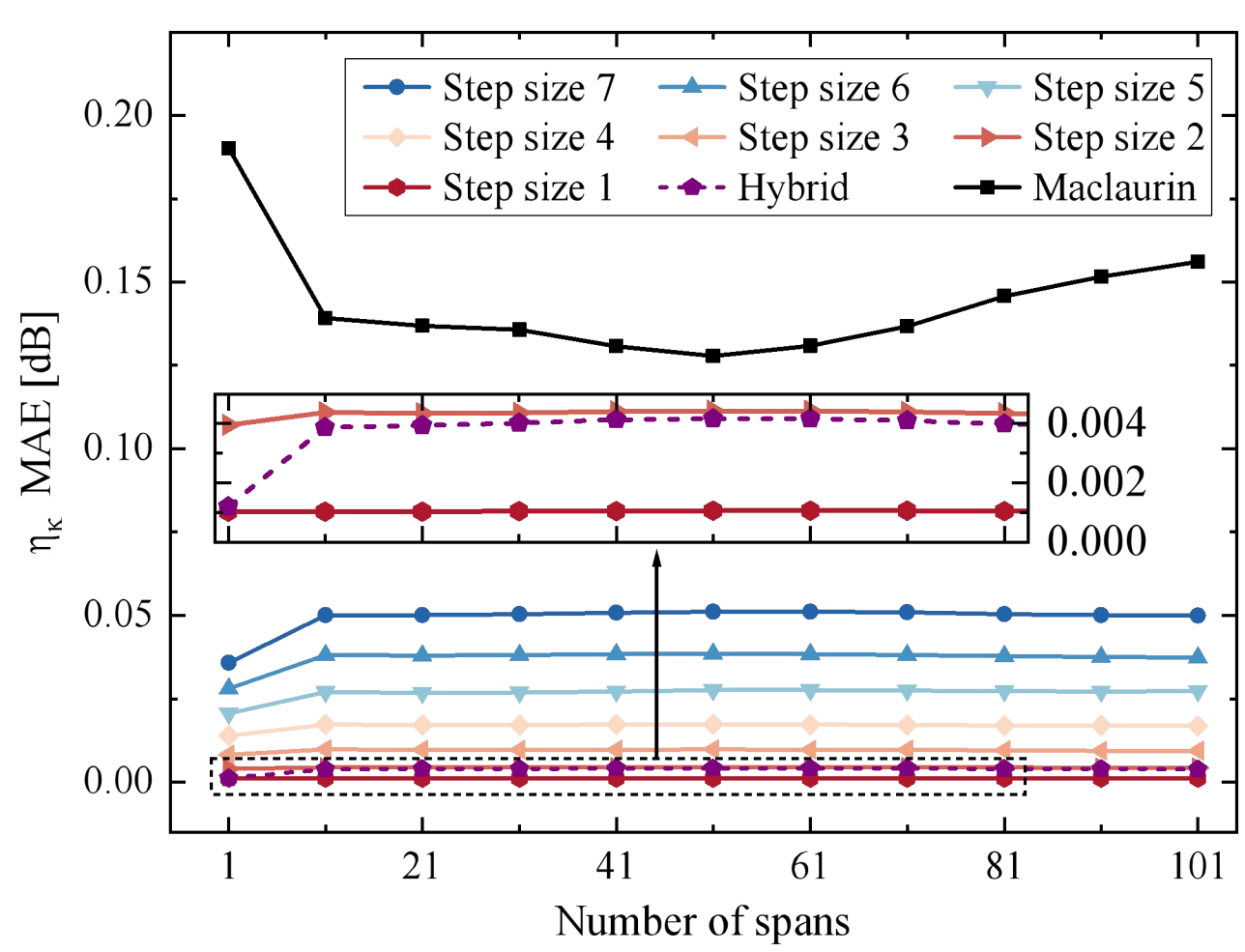

**Fig. 17.** Errors of various approximation methods when using the actual attenuation and Raman gain spectra.

## Acknowledgment

This work was supported in part by National Key Research and Development Program of China (2022YFB2903303) and National Natural Science Foundation of China (No. 62141505, 61367007 and 62371064).